\documentclass[reprint,twocolumn,superscriptaddress,showpacs,preprintnumbers,nofootinbib,aps,prd,
]{revtex4-1}

\usepackage{graphicx,color}
\usepackage{amsmath,amssymb,amsfonts}
\usepackage{stackrel}
\usepackage{enumitem}
\usepackage[english]{babel}
\usepackage{verbatim}
\usepackage{comment}
\usepackage{xcolor}
\usepackage{subcaption}
\usepackage{dcolumn}
\usepackage{physics}
\usepackage{units}
\usepackage{natbib}
\usepackage[section]{placeins}

\usepackage{eso-pic}

\usepackage[colorlinks=true,
            linkcolor=red,
            urlcolor=blue,
            citecolor=blue]{hyperref}
\graphicspath{
  {./}
  {figures/}
}

\newcommand{\tf}{\texorpdfstring}
\newcommand{\gev}{~\text{GeV}}
\newcommand{\tev}{~\text{TeV}}
\newcommand{\mev}{~\text{MeV}}

\newcommand{\onbb}{0\nu\beta\beta}

\def\nn{\nonumber}

\newcommand{\nldb}{0\nu\beta\beta} 

\definecolor{orange}{rgb}{1,0.5,0}
\definecolor{amethyst}{rgb}{0.6, 0.4, 0.8}
\definecolor{antiquefuchsia}{rgb}{0.57, 0.36, 0.51}
\definecolor{byzantine}{rgb}{0.74, 0.2, 0.64}
\definecolor{blue-violet}{rgb}{0.54, 0.17, 0.89}
\definecolor{cadmiumred}{rgb}{0.89, 0.0, 0.13}
\definecolor{brightcerulean}{rgb}{0.11, 0.67, 0.84}

\newcommand{\nus}{\nu_4} 
\newcommand{\mwr}{M_{W_R}}
\newcommand{\mzr}{M_{Z_R}}
\newcommand{\mnus}{m_{\nus}}

\newcommand{\mc}{\mathcal}

\newcommand{\uu}{\;\!}

\newcommand{\eg}{{\it e.g.}}
\newcommand{\ie}{{\it i.e.}}

\newcommand{\msq}{\lvert\mathcal{M}\rvert^2}
\newcommand{\rprod}{r_{\rm prod}}
\newcommand{\Gamabs}{\Gamma_{\rm sct}}
\newcommand{\Gamdec}{\Gamma_{\rm dec}}

\newcommand{\Renv}{R_{\rm env}}

\newcommand{\Rcore}{R_{\rm core}}

\begin{abstract}
The minimal left-right symmetric model (mLRSM) provides an elegant and testable 
framework for addressing the origin of neutrino masses. 
We examine the constraints on the sub-GeV right-handed (RH) neutrino in the type-II seesaw scenario of the mLRSM without left-right mixing, taking limits from collider searches, meson decays, supernovae, neutrinoless double beta ($\nldb$) decay and cosmology.
Specifically, we derive the $\nldb$ decay constraints using the advanced effective field theory approach and up-to-date nuclear matrix element calculations.
Besides, we update the SN1987A cooling bound with the state-of-the-art simulations, provide new constraints from the energy deposition in the supernova ejecta, and incorporate the stringent RH neutrino lifetime upper limit $\tau\lesssim 0.023\text{ s}$ from the big bang nucleosynthesis.
Our results identify the parameter region compatible with all current experimental and observational constraints, where the RH neutrino mass lies between 700~MeV and 1~GeV and the RH $W$ boson mass is slightly below 20~TeV.
This region is exclusively probed by the future tonne-scale $\nldb$ decay experiments,  providing a unique window to test the mLRSM and the possible origin of neutrino masses. 
\end{abstract}

\begin{document}

\setlength{\abovedisplayskip}{6pt}
\setlength{\belowdisplayskip}{6pt}

\title{ 
Sub-GeV Right-Handed Neutrino as a Probe of Neutrino Mass Generation in the Minimal Left-Right Symmetric Model
}
\author{Gang Li}
\email{ligang65@mail.sysu.edu.cn}
\affiliation{School of Physics and Astronomy, Sun Yat-sen University, Zhuhai 519082, China}
\affiliation{
Guangdong Provincial Key Laboratory of Quantum Metrology and Sensing, Sun Yat-Sen University, Zhuhai 519082, China}

\author{Ying-Ying Li}
\email{liyingying@ihep.ac.cn}
\affiliation{Institute of High Energy Physics, Chinese Academy of Sciences, Beijing 100049, China}

\author{Sida Lu}
\email{lusd5@mail.sysu.edu.cn}
\affiliation{School of Physics and Astronomy, Sun Yat-sen University, Zhuhai 519082, China}
\affiliation{
Guangdong Provincial Key Laboratory of Quantum Metrology and Sensing, Sun Yat-Sen University, Zhuhai 519082, China}

\author{Ye-Ling Zhou}
\email{zhouyeling@ucas.ac.cn}
\affiliation{School of Fundamental Physics and Mathematical Sciences, Hangzhou Institute for Advanced Study, UCAS, Hangzhou 310024, China}
\maketitle

\section{Introduction}
\label{sec:intro}

Neutrinos, among the most elusive particles, provide one of the clearest clues for physics beyond the standard model (BSM). 
The genesis of their unnaturally small masses remains a major 
puzzle despite longstanding endeavors. 
One of the most elegant approaches addressing this problem is the seesaw mechanism~\cite{Minkowski:1977sc, Yanagida:1979as, Gell-Mann:1979vob, Mohapatra:1979ia, Glashow:1979nm}.
With a Dirac mass of electroweak scale as a pivot, the masses of active neutrinos are naturally suppressed by the ultra-large right-handed (RH) Majorana neutrino masses. 
Yet, this most naive scenario expects the RH neutrinos to be as heavy as $10^{15}$ GeV, far beyond foreseeable experimental verifications. 
More testable solutions emerge when embedding the seesaw mechanism within left-right symmetric models~\cite{Pati:1974yy,Mohapatra:1974gc,Mohapatra:1974hk,Senjanovic:1975rk,Senjanovic:1978ev}. In particular, the minimal left-right symmetric model (mLRSM)~\cite{Mohapatra:1979ia,Mohapatra:1980yp} stands out for its notable predictability, and the possibilities of  parity restoration and grand unification at high energy scales~\cite{Senjanovic:2011zz}. 
Determining the origin of neutrino masses thus requires both verifying the possible Majorana nature of neutrinos and searching for additional RH particles, especially the RH neutrinos and gauge bosons.

As the most powerful particle generator ever, the Large Hadron Collider (LHC) can search for RH gauge bosons and neutrinos in the mLRSM thorough the well-known Keung-Senjanovi{\'c} process~\cite{Keung:1983uu}, setting a lower bound on the mass of the RH $W$ boson ($\mwr$) to be about 6 TeV in general~\cite{ATLAS:2023cjo,ATLAS:2019lsy,CMS:2022krd}.
Specifically, the same-sign lepton channel of the search can distinguish between Majorana and Dirac RH neutrinos.
Nevertheless, the sensitivity to $M_{W_R}$ does not improve much at the high-luminosity LHC, as the production cross section decreases significantly with the increase of $M_{W_R}$\footnote{Nevertheless, the sensitivity can be improved by considering displaced vertices searches in future far detectors. See, \eg, Ref.~\cite{deVries:2024mla}.}~\cite{Nemevsek:2018bbt}. 
Instead, in certain parameter region, more stringent constraints are provided by other experiments and observations. 
When an RH neutrino has a mass below GeV, which may appear as a ``sterile'' neutrino~\cite{Helo:2013esa}, it has been shown that future tonne-scale $\nldb$ experiments~\cite{LEGEND:2021bnm,nEXO:2021ujk,Agostini:2022zub} could be sensitive to $\mwr$ as large as 25 TeV~\cite{deVries:2022nyh}, which simultaneously provides the most sensitive and unambiguous probe of neutrinos' Majorana nature.
Various other constraints on $\mwr$ also exist for different sub-GeV RH neutrino mass ranges.
The SN1987A cooling bound has been estimated to exclude $\mwr\lesssim 23$ TeV for a RH neutrino mass smaller than 10 MeV~\cite{Barbieri:1988av}.
Meson decays have also been recently shown to exclude $\mwr \lesssim (4-19)$ TeV for RH neutrinos in the mass range $[50\mev, 1\gev]$~\cite{Alves:2023znq}. 
Other weaker constraints on $\mwr$ in the mLRSM can be found in Ref.~\cite{pdg_WR}.

In this work, we investigate and update the constraints on the sub-GeV RH neutrino in the type-II seesaw scenario of the mLRSM without left-right mixing.
For both the current $\nldb$ searches at KamLAND-Zen~\cite{KamLAND-Zen:2022tow,KamLAND-Zen:2024eml} and future tonne-scale experiments~\cite{LEGEND:2021bnm,nEXO:2021ujk,Agostini:2022zub}, we derive the constraints using the advanced effective field theory (EFT) approach and up-to-date nuclear matrix element (NME) calculations.
With the state-of-the-art simulations, we extend the applicability of the SN1987A cooling bound from RH neutrino mass below 10 MeV to a larger mass range, and provide a new constraint by considering the energy deposition in the supernova ejecta.
We also examined other constraints, including those from the LHC, meson decays and cosmology.
As a result, we identify parameter region that remains compatible with all current experimental and observational constraints and can be uniquely probed by future $\nldb$ decay experiments.

\section{Model setup }
\label{sec:model}

The mLRSM~\cite{Mohapatra:1979ia,Mohapatra:1980yp} is based on the gauge group $SU(2)_L \times SU(2)_R \times U(1)_{B-L}$, where $B$ and $L$ denote the baryon and lepton numbers, respectively.
Besides the Standard Model (SM) fields, three RH neutrinos, two scalar triplets, and an additional Higgs doublet are introduced, the last of which together with the SM Higgs furnishes a bidoublet of $SU(2)_L \times SU(2)_R$.  
More details of the model setup are given in Appendix~\ref{sec:mLRSM}. 

In the flavor basis, the charged current interactions are
\begin{align}
\label{eq:cc-interaction}
\mathcal{L}_{cc} = \dfrac{g}{\sqrt{2}} \left( \bar \ell_{L} \gamma^\mu \nu_{L} W_{L\mu}^{-} +  \bar \ell_{R} \gamma^\mu \nu_{R} W_{R\mu}^{-} \right) + {\rm h.c.}\;,
\end{align}
where h.c. denotes the Hermitian conjugate terms. 
The neutrino masses are described by
\begin{align}
\mathcal{L}_m = -\dfrac{1}{2} \left(\bar\nu_L,\bar\nu_R^c \right) 
\begin{pmatrix}
M_L & M_D \\
M_D^T & M_R
\end{pmatrix}
\begin{pmatrix}
\nu_L^c \\
\nu_R
\end{pmatrix}
+ {\rm h.c.}\;,
\end{align}
where 
the Majorana mass matrices are $M_L = \sqrt{2} Y_L^\dagger v_L$ and $M_R = \sqrt{2} Y_R v_R$, while the Dirac mass matrix is $M_D = (\kappa \Gamma_l  + \kappa^\prime \tilde{\Gamma}_l )/\sqrt{2}$.
Thus, the active neutrino masses are generated in both type-I and type-II seesaw mechanisms, with the mass matrix $M_\nu=M_L-M_DM_R^{-1}M_D^T$.

We will focus on the case without left-right mixing, 
namely both the mixing between $W_L$ and $W_R$ bosons and the mixing between LH and RH neutrinos (also known as active-sterile neutrino mixing) are ignored. 
A negligible $W_L$-$W_R$ mixing is reasonable in the limit of sufficiently small VEV ratio $\xi$ (cf. Eq.~\eqref{eq:LR-mixing}). The absence of active-sterile neutrino mixing occurs in the limit $M_D \to 0$, which implies that the masses of left-handed (LH) neutrino are generated via the type-II seesaw mechanism.

Moreover, we assume the generalized charge conjugation $(\mathcal{C})$ as the left-right symmetry~\cite{Maiezza:2010ic}\,\footnote{This parity is also called $D$-parity and is naturally predicted in $SO(10)$ grand unification framework~\cite{Chang:1983fu}.}.
Then, the Yukawa coupling matrices satisfy $Y_L = Y_R^\dagger$, so that $M_L = (v_L/v_R) M_R$. 
The LH and RH neutrino Majorana mass matrices can be diagonalized as follows:
\begin{align}
\widehat M_\nu = V_L^\dagger M_L V_L^*\;,\quad
\widehat M_N = V_R^T M_R V_R\;,
\end{align}
where $V_{L}$ is the Pontecorvo-Maki-Nakagawa-Sakata (PMNS) matrix, and $V_R$ is its RH analogue. It is natural to assume $V_R = V_{L}^*$, thus there is an one-to-one correspondence $\widehat M_\nu = (v_L/v_R) \widehat M_N$~\cite{Nemevsek:2012iq}.
The six neutrino mass eigenstates are labeled in ascending order by $\nu_i$, where $i=$1-3 correspond to LH neutrinos (collectively denoted $\nu$), whose masses are assumed in the normal hierarchy.
The absence of left-right mixing hence leads to a economical scenario of minimal free parameters $\{M_{W_R}, m_{\nu_4},  m_{\nu_6}\}$ (see more discussions in Appendix~\ref{sec:mLRSM}), among which $\mnus$ and $\mwr$ are of particular interest of this work.

\section{Experimental and observational constraints}
\label{sec:constraints}

From the searches for heavy gauge boson in the Keung-Senjanovi{\'c} process~\cite{ATLAS:2023cjo} and the final state of lepton plus missing energy~\cite{ATLAS:2019lsy,CMS:2022krd} at the LHC Run 2, we derive a lower bound $M_{W_R} \gtrsim 5.7\tev$ for $V_R = V_L^*$. 
For the low-energy processes, bounds from the measurements of $K$ and $B$ meson mixing exclude only $\mwr$ below around 3 TeV~\cite{Bertolini:2014sua,Bertolini:2019out} in the case of $\mathcal{C}$.
Meson decay bounds~\cite{Alves:2023znq} supersede the LHC ones in certain RH neutrino mass region, despite that in our case of RH neutrino mixing they are slightly weakened.
More discussions and other constraints from the LHC searches and meson decays are given in Appendix~\ref{app:collider}.
In the remainder of this section, we will investigate the constraints arising from $\onbb$ decay in KamLAND-Zen~\cite{KamLAND-Zen:2022tow,KamLAND-Zen:2024eml,Adams:2022jwx} and future tonne-scale experiments~\cite{LEGEND:2021bnm,nEXO:2021ujk,Agostini:2022zub},  as well as those from supernovae and cosmological observations.

\subsection{Neutrinoless double beta decay}
\label{sec:NLDBD}

We focus on $\onbb$ decay in the type-II seesaw scenario of the mLRSM~\cite{deVries:2022nyh,BhupalDev:2013ntw}. 
The inverse half-life of this process is expressed as~\cite{Cirigliano:2018yza}
\begin{align}
\left(T_{1/2}^{0\nu}\right)^{-1}&=g_A^4\left[G_{01}\left(|\mathcal{A}_L|^2+|\mathcal{A}_R|^2\right)\right.\nn\\
&\quad \left.-2(G_{01}-G_{04})\mathrm{Re}\mathcal{A}_L^*\mathcal{A}_R\right]\;,
\end{align}
where $g_A = 1.27$, $G_{01}=1.5\times 10^{-14}~{\rm yr}^{-1}$ and $G_{04}=1.2\times 10^{-14}~{\rm yr}^{-1}$ are phase-space factors for $^{136}\mathrm{Xe}$~\cite{Horoi:2017gmj}. Currently, the most stringent constraint on the half-life is $T_{1/2}^{0\nu} > 3.8\times 10^{26}~{\rm yr}$~\cite{KamLAND-Zen:2022tow,KamLAND-Zen:2024eml}, while future tonne-scale experiments are expected to reach the sensitivity $\sim 10^{28}~{\rm yr}$~\cite{nEXO:2021ujk,Agostini:2022zub}. In the EFT approach, the sub-amplitudes $\mathcal{A}_{L,R} = \sum_{i=1}^6 \mathcal{A}_{L,R} (m_{i})$, 
where $\mathcal{A}_{L}(m_{i})$ for $i\leq 3$ and $\mathcal{A}_{R}(m_{i})$ for $i>3$ are~\cite{Dekens:2020ttz,deVries:2022nyh}
\begin{align}
    \mathcal{A}_{L}(m_{i})&=-\frac{m_{i}}{4m_{e}}\left(\mathcal{M}_{V} +\mathcal{M}_{A} \right)\left(C_{\mathrm{VLL}}^{(6)}\right)_{ei}^{2}\nn\\
    &\qquad +\mathcal{A}_{L}^{(\nu)}(m_{i}) \;, \\
    \mathcal{A}_{R}(m_{i})&=-\frac{m_{i}}{4m_{e}}\left(\mathcal{M}_{V} +\mathcal{M}_{A} \right)\left(C_{\mathrm{VRR}}^{(6)}\right)_{e(i-3)}^{2}\nn\\
    &\qquad +\mathcal{A}_{R}^{(\nu)}(m_{i}) \;.
\end{align}
Here, $m_i\equiv m_{\nu_i}$ and $m_e$ denote the masses of neutrino $\nu_i$ and electron, respectively. 
The Wilson coefficients are
\begin{align}
    C_{\mathrm{VLL}}^{(6)}&=-2V_{ud} V_L\;,\quad
    C_{\mathrm{VRR}}^{(6)}=-\dfrac{v^{2}}{v_R^2} V_{ud} V_R \;,
\end{align}
which do not evolve between $M_{W_R}$ and $m_i$ for $m_i > \Lambda_\chi \simeq 1\gev $ under QCD running~\cite{deVries:2022nyh}.
The combinations of NMEs are expressed as
\begin{align}
    \mathcal{M}_{V}&=-\frac{g_{V}^{2}}{g_{A}^{2}}M_{F}(m_{i})+M_{GT}^{MM}(m_{i}) +M_{T}^{MM}(m_{i})\;,\\
    \mathcal{M}_{A}&=M_{GT}^{AA}(m_{i})+M_{GT}^{AP}(m_{i})+M_{GT}^{PP}(m_{i})\nn\\
    &\quad +M_{T}^{AP}(m_{i})+M_{T}^{PP}(m_{i})\;,
\end{align}
where $g_V=1$, $M_{F,GT,T}(m_i)$ denote the neutrino mass-dependent NMEs, which are obtained using the interpolation formulae given in Ref.~\cite{Dekens:2020ttz}. 
Note that sizable systematic uncertainties can arise from the many-body calculation methods of NMEs.
As discussed in Ref.~\cite{Ding:2024obt}, the NMEs for $^{136}\mathrm{Xe}$, including those corresponding to non-standard mechanisms, exhibit variations of approximately a factor of 2-4.

Compared with the traditional calculations of $\onbb$ decay half-life in the mLRSM~\cite{Doi:1985dx,Hirsch:1996qw,Vergados:2012xy}, the EFT approach requires the inclusion of the hard-neutrino exchange terms $\mathcal{A}_{L,R}^{(\nu)}$ in the sub-amplitudes, given by
\begin{align}
    \mathcal{A}_L^{(\nu)}(m_i)&=-\frac{m_i}{2m_e}\frac{m_\pi^2g_\nu^{NN}(m_i)}{g_A^2}\left(C_{\mathrm{VLL}}^{(6)}\right)_{ei}^2 M_{F,sd}\;,\\
    \mathcal{A}_R^{(\nu)}(m_i)&=-\frac{m_i}{2m_e}\frac{m_\pi^2g_\nu^{NN}(m_i)}{g_A^2}\left(C_{\mathrm{VRR}}^{(6)}\right)_{e(i-3)}^2 M_{F,sd}\;,
\end{align}
where the low-energy constant (LEC) is~\cite{Dekens:2020ttz,Dekens:2024hlz}
\begin{align}
    g_\nu^{NN}(m_i)&=g_\nu^{NN}(0) \dfrac{1 + (m_i/\Lambda_\chi)^2}{1+(m_i/\Lambda_\chi)^2 (m_i/m_\pi)^2}\;,
\end{align}
with the LEC for the exchange of LH neutrinos is $g_\nu^{NN}(0) = -(1.055\pm 0.385)~\mathrm{fm}^2$~\cite{Jokiniemi:2021qqv}. 
These contact terms are at the leading order~\cite{Cirigliano:2018hja, Cirigliano:2019vdj,Cirigliano:2020dmx,Cirigliano:2021qko} in Weinberg power counting~\cite{Weinberg:1990rz,Weinberg:1991um}, which significantly enhance the $\onbb$ decay amplitude\,\footnote{In the standard mechanism of $\onbb$ decay, such an enhancement was obtained in Ref.~\cite{Wirth:2021pij}.}. 
We adopt the intermediate value for the LEC $g_\nu^{NN}(0)$~\cite{Jokiniemi:2021qqv,Dekens:2024hlz}, such that the uncertainties in our calculations are mainly sourced by the NMEs~\cite{Menendez:2017fdf,Hyvarinen:2015bda,Graf:2022lhj,Ding:2024obt}.

\subsection{Supernova constraints}
\label{sec:sn-constraint}

Supernovae (SNe) are ideal test grounds for the sub-GeV RH neutrino and the underlying mLRSM, as their explosion dynamics and relevant observations may be significantly influenced by the light BSM particles.
Below we briefly explain how the SN constraints are drawn, leaving more calculation details in Appendix~\ref{app:SN_constraints}.

The sub-GeV RH neutrino $\nus$s are mainly produced in the core of the SN via $e^-+p \to \nus+n$\,\footnote{The abundance of muons~\cite{Bollig:2020xdr} in SN is usually less than that of electrons, especially during the first second of the explosion where the signals are mainly generated. The muonic counterpart of this production channel is thus ignored.}.
Note that the leading production channel in the type-I seesaw scenario, $\nu+N\to \nus+N$~\cite{Carenza:2023old,Chauhan:2025mnn}, is absent here due to the vanishing Dirac mass $M_D$. 
The other possible channel, $\nu+\nu\to\nus+\nus$ mediated by the $Z_R$ boson, turns out to be sub-leading.
For our calculations, we use the simulation SFHo-18.6 in Ref.~\cite{Bollig:2020xdr} performed by the Garching group for the profiles of all necessary thermodynamics quantities. 

As $\nus$s propagate to the outside of the star through the envelope region after being produced, their flux is modified for various reasons.
Their energy inevitably experiences a redshift induced by the stellar body's gravitational potential.
More importantly, their flux intensity is dissipated during propagation, due to their own decay, the stellar body's gravitational trapping or the stellar envelope's absorption through the scatterings $\nus+n\to e^- + p$ and $\nus+p\to e^++n$.
These fail-to-escape $\nus$ can deposit energy into the stellar envelope, increasing the total {\it explosion} energy (\ie, the kinetic energy of the ejecta in the explosion)~\cite{Fiorillo:2025yzf,Caputo:2022mah,Sung:2019xie,Falk:1978kf}.
The constraint on the total deposited energy is usually taken to be $E_{\rm depo}\lesssim 10^{50}\text{ erg}$~\cite{Carenza:2023old}, about 10\% of the total explosion energy.
$e^\pm$ produced directly or secondarily (pions and muons are assumed as decaying promptly)  from the possible decay of $\nus$ is always a viable energy deposition channel, and the LH neutrinos have also been shown to be efficiently absorbed when the ambient density is greater than about $10^{12}\,\text{g/cm}^3$~\cite{Rembiasz:2018lok}.
Furthermore, in principle the two aforementioned scattering processes should also contribute to the energy deposition, as the final products are easily trapped in the ejecta.
Note that the scattering absorption processes were not considered when SNe were used to constrain light BSM particles like dark photons or axion-like particles~\cite{Fiorillo:2025yzf,Caputo:2022mah,Sung:2019xie}. 

The $\nus$s that escape the SN carry away energy, contributing an additional cooling channel for the star. 
With the total budget of {\it released} energy of an SN fixed to be about $10^{53}$ erg, the LH neutrino flux from an SN will be suppressed if a non-negligible portion of the energy is emitted via $\nus$, which leads to the so-called cooling bound.
To have the neutrino flux from SN1987A to be compatible with experimental observations~\cite{Kamiokande-II:1987idp,Bionta:1987qt,Alekseev:1988gp}, the cooling bound requires the luminosity of the new emission channels to be less than $3\times 10^{52}\text{ erg/s}$~\cite{raffelt1996stars}, which is roughly equivalent to requiring the new particles to take less than about 10\% of SN1987A's total released energy~\cite{Dreiner:2003wh}.

\subsection{Cosmological constraints} 
\label{sec:cosmo_constraints}

Stringent constraints on sub-GeV RH neutrinos also arise from the modifications brought to the several crucial cosmic evolution eras, specifically the BBN and the cosmic microwave background (CMB).
Due to the relatively large interaction rate, the $\nus$ in the mLRSM will inevitably thermalize during the early universe and then freeze out, see Appendix~\ref{app:nus_thermal_history} for more details on $\nus$'s thermal history.
Meanwhile, the decay products of $\nus$ may spoil the abundance of protons, neutrons or light elements like $^4$He from the BBN through extra entropy injection, changing the cosmic expansion rate, or shifting the equilibrium of the reactions.
The sub-GeV RH neutrinos are thus expected to decay either much earlier or much later than the time period relevant for the BBN, the former of which translates to an {\it upper} bound on $\mwr$.
For $\mnus>m_{\pi}$ where the hadronic decay of $\nus$ is available, the most stringent constraint is obtained from the injected pions' shifting the $p\leftrightarrow n$ conversion equilibrium, requiring the lifetime of $\nus$ to be $\tau_{\nus}\lesssim 0.023\text{ s}$\footnote{Despite obtained from a 3+1 active-sterile neutrino mixing model, this lifetime constraint can be directly translated to the mLRSM, as both models are dominated by the same hadronic decay mode in this parameter region.
It is worth noting that a weaker condition $\tau_{\nus}\lesssim 1\text{ s}$ was employed in Refs.~\cite{Nemevsek:2011aa, Nemevsek:2023hwx}.}~\cite{Boyarsky:2020dzc,Chen:2024cla}.
Though this BBN constraint becomes irrelevant for $\tau_{\nus}\gtrsim 10^4\text{ s}$~\cite{Ovchynnikov:2021zyo}, that $\nus$'s abundance after its freeze-out is much larger than the dark matter relic abundance measured from the CMB~\cite{Planck:2018vyg} implies the parameter space to be still ruled out either by $\nus$'s relic abundance (for long-lived $\nus$) or $\Delta N_{\rm eff}$ at CMB (for $\nus$ decaying before CMB, which injects additional relativistic degrees of freedom to the plasma).

For $m_{\nus}<m_\pi$, on the other hand, the majority of the existing BBN constraints on sub-GeV RH neutrinos (\eg, Ref.~\cite{Sabti:2020yrt}) doesn't apply for the mLRSM, since they are derived with the 3+1 mixing model where sterile neutrinos decay mainly into $l^+l^-\nu$ and $3\nu$~\cite{Bondarenko:2018ptm}, while the $\nus$s are stable in this mass range for the mLRSM setup of our interest.
Despite no energy injection is made to the plasma, the excessive abundance of $\nus$ from their freeze-out can still spoil either the BBN or the CMB, and thus rules out the entire parameter space.

From the model-building point of view, one possible approach to evade the cosmological constraint in this region is to introduce an additional completely invisible decay channel for $\nus$ to efficiently deplete its abundance before BBN~\cite{Brdar:2020quo}, while the details are beyond the scope of the mLRSM and hence the topic of this work.
But if such possibility is feasible, this invisible decay channel is expected to strengthen the SN cooling bound and weaken the energy deposition constraint, as the decay products could still carry away energy from the SN but are exempted from being absorbed by the stellar envelope.

\section{Results and discussions} 
\label{sec:result}

\begin{figure}[!htb]
\centering
\includegraphics[scale=0.7]{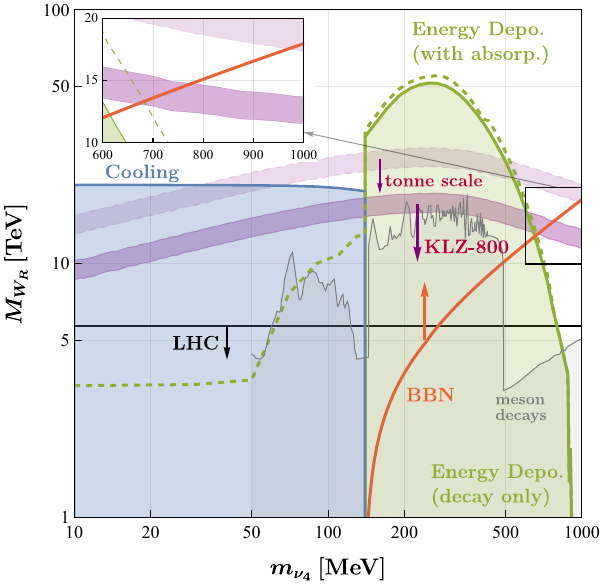}
\caption{Constraints on $\mwr$ from the SN1987A cooling (blue), SN energy deposition (green), $\nldb$ decay (purple), BBN (red), meson decays (gray, taken from Ref.~\cite{Alves:2023znq}) and LHC searches (black) for various $\mnus$.
Two variations are considered for the SN energy deposition constraint, one of which (solid green) ignores the $\nus$ scattering absorption among the sources, while the other (dashed green) has it included. 
The top-left inset panel zooms into the region enclosed by the black square.
}
\label{fig:exclusion}
\end{figure}

In Fig.~\ref{fig:exclusion}, we summarize all the discussed constraints in the previous section for $\mnus<1\text{ GeV}$ on the $\mnus$-$\mwr$ plane for $V_R = V_L^*$ in the type-II seesaw scenario without $W_L$-$W_R$ mixing (see Appendix~\ref{app:LR-mixing} for results with the mixing). 
The lower limit $M_{W_R} = 5.7\tev$ from the LHC searches~\cite{ATLAS:2019lsy,CMS:2022krd,ATLAS:2023cjo} is depicted as the black line. The $\nldb$ decay constraints from KamLAND-Zen~\cite{KamLAND-Zen:2024eml} and the projected sensitivity of future tonne-scale experiments~\cite{nEXO:2021ujk,Agostini:2022zub} are shown as purple bands, where the band width reflects the uncertainties in NME calculations.
With $m_{\nu_4}$ lying between 100~MeV and 1~GeV and $m_{\nu_6}$ around the TeV scale, the $\onbb$ decay rate is mainly contributed by $\nus$ for $M_{W_R} \lesssim 25\tev$~\cite{deVries:2022nyh} and hence directly constrains the sub-GeV RH neutrino.
Note that our derived $\mwr$ constraint from current $\nldb$ decay experiments is stronger than that from meson decays, different from~\cite{Alves:2023znq}, which is attributed to the more advanced EFT approach~\cite{Cirigliano:2018yza,Dekens:2020ttz,deVries:2022nyh} as well as the more recent NME calculations adopted in this work.
In comparison, the $\onbb$ decay result in Ref.~\cite{Alves:2023znq} is based on an 
old-valued NME calculated with QRPA method~\cite{Pantis:1996py}.

For the SN constraints, SN1987A's cooling poses a nearly flat lower bound on $\mwr\sim 20$ TeV, comparable to that obtained in Ref.~\cite{Barbieri:1988av} in the mass range of $m_{\nu_4} \lesssim 10$MeV\,\footnote{
Ref.~\cite{Barbieri:1988av} derived a lower bound of $\mwr\gtrsim 23\text{ TeV}$, assuming no $W_L$-$W_R$ mixing and  $|(V_R)_{e1}| = 1$, also neglecting the mass of $\nu_4$. As pointed out in Ref.~\cite{Nemevsek:2011aa}, when RH neutrino mixing is taken into account, this limit generalizes to $\mwr\gtrsim \sqrt{|(V_R)_{e1}|}\times 23\text{ TeV}$. For $V_R = V_L^*$, the resulting lower bound is approximately $21\tev$~\cite{deVries:2022nyh}. We note that other constraints from the SN energy deposition, BBN and $\onbb$ decay depend on $V_R$ in the same manner as the SN cooling bound, and the relative positions between the constraints thus do not change with respect to $V_R$.
}. 
Note that our result extends existing SN1987A cooling bound to beyond $\mnus=10$ MeV in the mLRSM.
The bound terminates abruptly at $\mnus= m_{\pi}$, corresponding to the turn-on of $\nus$'s hadronic decay which greatly increases the dissipation of the $\nus$ flux in the SN envelope.
For the energy deposition in the SN envelope, two different variants of the constraint are presented.
One of them (the solid green contour, marked as ``decay only'') includes only $e^\pm$  and $\nu$ 
(included conditionally, see discussions in Sec.~\ref{sec:sn-constraint}) from $\nus$'s decay as the sources of injection, to match the processes considered in earlier works on axion-like particles and dark photons~\cite{Caputo:2022mah,Sung:2019xie}.
The other (dashed green, marked as ``with absorption'') also includes the $\nus$ scattering processes $\nus+n\to e^-+p$ and $\nus+p\to e^++n$, which increases the energy deposition rate and thereby provide stronger constraints.
The conservative result starts at the $\mnus= m_{\pi}$ for exactly the same reason as the termination of the cooling bound, while the one with absorption included can extend to a lower mass region.

Despite SN in general provides stronger constraints on $\mwr$ than the $\nldb$ decay constraints for $\mnus\lesssim 600$ MeV, the cosmological constraint turns out to be much more powerful and excludes nearly all the parameter space. 
Beyond the SN constraints, the $\nldb$ decay constraint from KamLAND-Zen together with the cosmological constraint leaves a window for the mLRSM at $\mnus\gtrsim 700\text{ MeV}$ and $\mwr\lesssim 20\text{ TeV}$, which lies within the probing capability of the next-generation tonne-scale $\nldb$ decay experiments exactly. 


Nevertheless, that SNe as booming test grounds of BSM physics is nourishing both new phenomenological observables and the fundamental understanding of SN itself.
Several other possible observables have been recently proposed, including fireball formation, secondary neutrinos or SN boosted dark matter~\cite{Diamond:2023cto,Fiorillo:2022cdq,Akita:2022etk,Liang:2024xcx}.
Some of them could potentially be stronger than the cooling and energy deposition bound.
Also, the current SN simulations are shown to be not in full consistency with the sparse SN1987A data~\cite{Li:2023ulf,Fiorillo:2023frv}, and more sophisticated simulations (\eg, Refs.~\cite{Bollig:2020phc,Nakamura:2022zlc} or full 3D simulations) may be necessary.
It would be interesting to see what all these would imply for the scenario of this work and the SN constraints on BSM physics in general.

\begin{acknowledgments}
We would like to thank Michael J. Ramsey-Musolf, Edoardo Vitagliano and Damiano F.G. Fiorillo for helpful discussions, and Hans Thomas-Janka and Daniel Kresse for providing access to the SN simulation data used in this work~\cite{Bollig:2020xdr}.
This work is supported by the National Science Foundation of China under Grant Nos.~12347105, 12305107, 12205064, 12347103.
GL is also supported by the Guangdong Basic and Applied Basic Research Foundation (2024A1515012668), and SYSU startup funding.
YLZ is also supported by Zhejiang Provincial Natural Science Foundation of China under Grant No. LDQ24A050002.

\end{acknowledgments}

\onecolumngrid
\appendix

\section{Minimal left-right symmetric model}
\label{sec:mLRSM}

In the mLRSM~\cite{Mohapatra:1979ia,Mohapatra:1980yp}, the lepton doublets are defined as
\begin{align}
L_{L}=\left(\begin{array}{l}
\nu_L \\
\ell_L
\end{array}\right)\;, \quad L_{R}=\left(\begin{array}{c}
\nu_R \\
\ell_R
\end{array}\right)\;.
\end{align}

The Yukawa interactions are given by
\begin{align}
\mathcal{L}_Y  &=  -\bar{Q}_L (\Gamma \Phi + \tilde{\Gamma} \tilde{\Phi}) Q_R  
-\bar{L}_L (\Gamma_l \Phi + \tilde{\Gamma}_l \tilde{\Phi}) L_R  \nn\\
&\quad - \left( \bar{L}_L^c i\tau_2 \Delta_L Y_L L_L +  \bar{L}_R^c i\tau_2 \Delta_R Y_R L_R\right) + {\rm h.c.}\;.
\end{align}
Here, $\Delta_{L,R}$ are scalar triplets, and $\Phi$ is the scalar bidoublet with $\tilde \Phi  \equiv \tau_2 \Phi^* \tau_2$, where $\tau_2$ is the second Pauli matrix.
$\Gamma$, $\tilde{\Gamma}$, $\Gamma_l$ and $\tilde{\Gamma}_l$ represent the Yukawa coupling matrices, and $\bar L_{L,R}^c \equiv L_{L,R}^T C$ with $C$ being the charge conjugation matrix.

After the symmetry breaking, the scalar fields obtain vacuum expectation values (VEVs) as follows:
\begin{align}
\label{eq:triplets}
\left\langle\Delta_L\right\rangle=\left(\begin{array}{ccc}
0 & 0 \\
v_L e^{i \theta_L} / \sqrt{2} & 0
\end{array}\right), \ 
\left\langle\Delta_R\right\rangle=\left(\begin{array}{cc}
0 & 0 \\
v_R / \sqrt{2} & 0
\end{array}\right)
\end{align}
and 
\begin{align}
\label{eq:bidoublet}
\langle\Phi\rangle=\left(\begin{array}{cc}
\kappa/\sqrt{2} & 0 \\
0 & \kappa^\prime e^{i \alpha}/\sqrt{2}
\end{array}\right)\;.
\end{align}
where the VEVs $v_L$, $v_R$, $\kappa$ and $\kappa^\prime$ are real~\cite{Deshpande:1990ip,Zhang:2007da}. The phases $\alpha$ and $\theta_L$ can induce $CP$ violation, which will be set as zero throughout this work.  
The masses of the charged gauge bosons $W_L$ and $W_R$ are
\begin{align}
M_{W_{L,R}}^2 &= \dfrac{g_{L,R}^2}{4} \left(\kappa^2 + \kappa^{\prime 2} + 2 v_{L,R}^2\right) \;,
\end{align}
where we have assumed the gauge couplings of $SU(2)_R$ and $SU(2)_L$ are equal, $g_R = g_L \equiv g \simeq 0.65$, ignoring the small renormalization-goup evolution effect. 
The measurements of electroweak $\rho$ parameter indicates that the triplet VEV $v_L \ll v \equiv \sqrt{\kappa^2 + \kappa^{\prime 2}}=246\gev$~\cite{Deshpande:1990ip}.

In the mass basis, the mixing between $W_L$ and $W_R$ is described as
\begin{align}
\begin{pmatrix}
W_L^{\pm} \\
W_R^{\pm}
\end{pmatrix} = 
\begin{pmatrix}
\cos\zeta & -\sin\zeta \\
\sin\zeta & \cos\zeta
\end{pmatrix}
\begin{pmatrix}
W_1^{\pm}\\
W_2^{\pm}
\end{pmatrix}\;,
\end{align}
where $W_1$ and $W_2$ are the mass eigenstates. The mixing parameter $\zeta$ is given by
\begin{align}
\label{eq:LR-mixing}
\tan\zeta = \lambda \dfrac{2 \xi}{1+ \xi^2} \;, \quad
\xi \equiv \kappa^\prime/\kappa\;,\quad
\lambda \equiv M_{W_L}^2/\mwr^2\;.
\end{align}

The $W_L$-$W_R$ mixing has drawn lots of attention. It may give leading contribution to the $\onbb$ decay rate~\cite{Li:2020flq}, and has important implications for neutron electric dipole moments  and  kaon $CP$ violation~\cite{Bertolini:2019out,Ramsey-Musolf:2020ndm,Dekens:2021bro}.
We set the phases $\alpha$ and $\theta_L$ to be zero, so these constraints are evaded~\cite{Bertolini:2019out}.
The most stringent experimental constraint on the mixing parameter $\zeta$ arises from measurements of the electroweak $\rho$ parameter~\cite{Czakon:1999ga,Li:2022cuq,deVries:2022nyh}.  
For $\mwr = 7\tev$, we obtain an upper bound on the VEV ratio $\xi \lesssim 0.6$ using Eq.~\eqref{eq:LR-mixing} and the latest global fit result $\rho = 1.00031\pm 0.00019$~\cite{ParticleDataGroup:2024cfk}. Besides, there is a theoretical bound from the perturbativity of Yukawa couplings, requiring $\xi < 0.77$~\cite{Maiezza:2010ic,Nemevsek:2023yjl}. 

We define the physical masses of neutrinos as $\widehat M_\nu = {\rm diag}\{m_{\nu_1}\,, m_{\nu_2}\,, m_{\nu_3}\}$ and $\widehat M_N = {\rm diag}\{m_{\nu_4}\,, m_{\nu_5}\,, m_{\nu_6}\}$.
In the type-II seesaw scenario of the mLRSM with $V_R = V_L^*$, if the LH neutrino masses are in the normal hierarchy, we have
\begin{align}
  m_{\nu_{4,5}} = \dfrac{m_{\nu_{1,2}}}{m_{\nu_3}} m_{\nu_6}\;.
\end{align}
By using the measurements of the neutrino masses, which are sensitive to the squared mass difference, we find that masses of RH neutrinos $\nu_4$ and $\nu_5$ are determined by the lightest LH neutrino mass $m_{\nu_1}$ and the heaviest RH neutrino mass $m_{\nu_6}$. Numerically, we obtain~\cite{deVries:2022nyh}
\begin{align}
\label{eq:RHmasses}
    m_{\nu_4} \simeq 20\mev \cdot \dfrac{m_{\nu_6}}{1\tev} \cdot \dfrac{m_{\nu_1}}{10^{-6}~{\rm eV}}\;,\quad
    m_{\nu_5} \simeq 175\gev \cdot \dfrac{m_{\nu_6}}{1\tev}\;.
\end{align}
Thus, if $m_{\nu_6} = 1~\text{TeV}$, the mass of $\nu_4$ can vary between $10 ~\text{MeV}$ and $1 ~\text{GeV}$ for $5\times 10^{-2}<\frac{m_{\nu_1}}{10^{-5}~\mathrm{eV}}  \lesssim 5 $, which satisfies the upper limit on the absolute neutrino masses $m_\nu < 0.45~{\rm eV}$ set by the KATRIN experiment~\cite{KATRIN:2024cdt}, as well as the bound on the sum of neutrino masses $\sum_{i=1}^3 m_{\nu_i} < 0.064~{\rm eV}$ from DESI~\cite{DESI:2025zgx}.
In this work, we take $\{M_{W_R}, m_{\nu_4},  m_{\nu_6}\}$ as the input parameters, following the approach suggested in Ref.~\cite{Tello:2010am} and adopted in Refs.~\cite{Li:2020flq,deVries:2022nyh}.
Assuming $m_{\nu_6}$ is at the TeV scale, we find that its precise value has little impact on our results. This is because the heavier two RH neutrinos give much smaller contributions to the $\onbb$ decay rate compared to $\nu_4$~\cite{deVries:2022nyh}. Besides, the best lower limit on $M_{W_R}$ from searches at the LHC Run 2 remains unchanged, as discussed in Appendix~\ref{app:collider}.

\section{Constraints from LHC searches and meson decays}
\label{app:collider}

In this appendix, we will investigate the constraints on $M_{W_R}$ for sub-GeV RH neutrino in the mLRSM for $V_R = V_L^*$ from the searches at the LHC Run 2, and discuss those from meson decays.

To derive the constraints from searches for heavy charged boson in the final state of lepton plus missing energy~\cite{ATLAS:2019lsy,CMS:2022krd}, we can simply multiply the benchmark theoretical signal cross sections by $|(V_R)_{\alpha 1}|^2$ (where $\alpha = e, \mu$) in the electron and muon channels, respectively. We then obtain the lower limit of $ M_{W_R} \gtrsim 5.7~\text{TeV}$ for $V_R = V_L^*$. 
For the searches in the Keung-Senjanovi{\'c} process~\cite{ATLAS:2023cjo}, the exclusion limits are presented in the plane of the heavy RH neutrino mass $(\gtrsim 87.5\gev)$ and $M_{W_R}$. The most stringent constraint on $M_{W_R} \gtrsim 6.4\tev$ occurs for the heavy RH neutrino mass being about $1\tev$.
In the scenario of the mLRSM we consider, the masses of RH neutrino are correlated (cf. Eq.~\eqref{eq:RHmasses}), we can obtain that the most stringent constraint on $M_{W_R} \gtrsim 5.7\tev$. In practice, we assume the RH neutrino mass being 1~TeV, and solve the relation $\sigma(M_{W_R}) \times |(V_R)_{\alpha i}|^2 = \sigma(M_{W_R} = 6.4\tev)$ for $\alpha = e, \mu$ and $i=2,3$ to derive the limit on $M_{W_{R}}$, where $\sigma(M_{W_R})$ is the theoretical cross section of the signal process for varying $M_{W_R}$ as given in Ref.~\cite{Nemevsek:2018bbt}. Other choices of RH neutrino masses result in weaker constraints on $M_{W_{R}}$.

There are also constraints from the LHC searches for the neutral gauge boson $Z_R$~\cite{ATLAS:2019erb,CMS:2021ctt}. From recent recast for $g_R = g_L$, the lower limit on the $Z_R$ mass is $M_{Z_R} > 5.36\tev$~\cite{Lichtenstein:2025pxs}. In the mLRSM, the masses of $W_R$ and $Z_R$ are correlated as $M_{Z_R} \simeq 1.67\, M_{W_R}$~\cite{Kriewald:2024cgr}, implying that constraints on the mLRSM from the $Z_R$ searches are not as stringent as those from the $W_R$ searches~\cite{ATLAS:2023cjo}.

Recently, Ref.~\cite{Alves:2023znq} conducted a systematic study of the constraints on the mass of $W_R$, deriving a lower bound of $M_{W_R} \gtrsim (4-19) \tev$ for RH neutrinos in the mass range $[50 \mev, 1.9 \gev]$. 
Their model setup is similar to ours, with the exception that they assumed the masses of the three RH neutrinos to be degenerate. This assumption effectively set $(V_R)_{\alpha 1} =1$ in the decay widths of meson into RH neutrino. 
If $V_R = V_L^*$ is considered, the constraints on $M_{W_R}$ from meson decays would be slightly relaxed by a factor of $1/\sqrt{|(V_L)_{\alpha 1}|}$ with $\alpha = e, \mu$, which is a factor of $1.1$ or $1.66$ in the electron and muon channels, respectively.

\section{Details on the supernova constraints on the sub-GeV RH neutrinos}
\label{app:SN_constraints}

In this Appendix we present the details on our calculations of the SN constraints in Sec.~\ref{sec:sn-constraint}.
Specifically, we will explain how the sub-GeV RH neutrino production and how their propagation in the stellar envelope are treated, and how the constraints are obtained.

\subsection{Production}

The production rate of $\nus$ per unit volume and energy is given by~\cite{Carenza:2023old}
\begin{align}
\dfrac{d^2n_{\nus}}{dE_{\nus} dt}\approx\int&\dfrac{d^3p_e}{(2\pi)^3(2E_e)}\dfrac{d^3p_p}{(2\pi)^3(2E_p)}\dfrac{d^3p_n}{(2\pi)^3(2E_n)}\dfrac{p_{\nus}}{4\pi^2}(2\pi)^4\delta^{(4)}(p_e+p_p-p_{\nus}-p_n)\cdot \msq f_{e}f_p(1-f_n)\,,\label{eq:d2ndEdt}
\end{align}
where
\begin{equation}
\begin{aligned}
\msq=&\,\dfrac{M^4_{W_L}G^2_F\lvert V_{ud}\rvert^2}{4\mwr^4m_N^2}\Big[16 F_1 t G_A m_N^2 \left(m_{\nu _4}^2+2 m_N^2-t-2 u\right)+16 G_A G_P m_{\nu _4}^2 m_N^2 \left(t-m_{\nu _4}^2\right)\\
&\,+8 G_A^2 m_N^2 \left(-4 m_N^2 \left(-m_{\nu _4}^2+t+u\right)+2 m_N^4-m_{\nu _4}^2 (t+2 u)+t^2+2 t u+2 u^2\right)\\
&\,+8 F_1^2 m_N^2 \left(-4 u m_N^2+2 m_N^4-m_{\nu _4}^2 (t+2 u)+t^2+2 t u+2 u^2\right)+4t G_P^2 m_{\nu _4}^2 \left(t-m_{\nu _4}^2\right)\\
&\,+8 F_2 m_N^2 \left(2 t G_A \left(m_{\nu _4}^2+2 m_N^2-t-2 u\right)+F_1 \left(-m_{\nu _4}^4-t m_{\nu _4}^2+2 t^2\right)\right)\\
&\,-F_2^2 \left(4 m_N^2 \left(m_{\nu _4}^4+t m_{\nu _4}^2-2 t (t+u)\right)+4 t m_N^4+t \left(m_{\nu _4}^4-m_{\nu _4}^2 (t+4 u)+4 u (t+u)\right)\right)\Big]\,,
\end{aligned}
\end{equation}
is the matrix element of the production process~\cite{Barbieri:1988av,Giunti:2007ry}, $t$ and $u$ are Mandelstam variables,
and $f_i(p_i)$ is the momentum distribution of species $i$. We have assumed $f_{\nus}\ll 1$ such that the back-reactions can be ignored.
With the involvement of nucleons in the process, the matrix element $\msq$ contains the nucleon form factors $F_1$, $F_2$, $G_A$ and $G_P$ which are functions of squared transferred momentum $q^2$~\cite{Giunti:2007ry,Guo:2020tgx}.
Since these form factors involve quantities that are typically around GeV and are much larger than the typical transferred momentum, their $q^2$-dependence is therefore ignored in our calculation\footnote{In fact, $G_P$ also contains $m_\pi$ which could be comparable to the transferred momentum. We have checked that turning on the $q^2$-dependence in $G_P$ would lead to $\mc{O}(10\%)$ difference in the calculation, and hence ignore this $q^2$-dependence as well.}.
The formidable dimension-9 integration can then be reduced to a dimension-2 one~\cite{Hannestad:1995rs}, greatly benefiting the numeric evaluations.
Note that nucleons may have their self-energy significantly modified in dense environments, which corrects their effective masses and chemical potentials in $\msq$ and $f_N$~\cite{Hempel:2014ssa}.

As mentioned in the main text, $\nus$s are produced mainly in the core of the SN. 
We therefore consider the production only within the radius $\Rcore=40$ km.

\subsection{Propagation and dissipation in the stellar envelop}

As explained in Sec.~\ref{sec:sn-constraint}, the $\nus$ experiences a gravitational redshift and a dissipation in flux intensity caused by $\nus$'s decay and environmental absorption as $\nus$ propagating towards the outside of the SN.
The gravitational redshift is accounted for by a ``lapse factor'' $\alpha(r)$, which is provided in the SFHo-18.6 simulation data~\cite{Bollig:2020xdr}.
A $\nus$ with energy $E$ at radius $r$ has its energy at radius $r^\prime$ to be $E^\prime=\alpha(r)E/\alpha(r^\prime)$.
Assuming a radial motion for the produced $\nus$s, the flux dissipation for $\nus$s produced at radius $r$ with energy $E$ is characterized by their survivability at $r^\prime$ as
\begin{align}
P_{\rm srv}(E,r;\uu r^\prime)=\exp\left[-\int^{r^\prime}_{r} \dfrac{dl}{\beta_{\nus}\big(E^\prime(l)\big)}\ \Gamma(E^\prime(l))\right]\Theta(E^\prime-\mnus)\uu,\label{eq:P_srv}
\end{align}
where $\beta_{\nus}(E^\prime)$ is the velocity of a $\nus$ with energy $E^\prime$, $\Theta(z)$ is the Heaviside function included to account for the gravitational trapping, and $\Gamma$ it the total loss rate which is the summation of the decay rate $\Gamdec$ and scattering absorption rate $\Gamabs$
\begin{align}
\Gamma(E^\prime)=\Gamabs(E^\prime)+\Gamdec(E^\prime)\,.
\end{align}

Given the charged-current interactions in Eq.~\eqref{eq:cc-interaction} and the absence of left-right mixing, a $\nu_4$ can only decay into the hadronic final state as $\nu_4 \to \ell^\mp_j \pi^\pm $,
whose width at $\nu_4$'s rest frame is given by~\cite{Bondarenko:2018ptm} 
\begin{align}\label{eq:Gamma_had}
\Gamma_{\rm dec,0}&= \dfrac{G_F^2 f_\pi^2 m_{\nu_4}^3 }{8\pi } |V_{ud}|^2 \lambda^2 \sum_{\alpha = e,\mu} \theta ( 1-x_\pi -x_\alpha)  \left| (V_R)_{\alpha 1} \right|^2 \left[\left(1-x_\alpha^2\right)^2-x_\pi^2(1+x_\alpha^2)\right]  \beta(1,x_\pi^2,x_\alpha^2)\,,  
\end{align}
where 
$V_{ud}$ denotes the $(1, 1)$ element of the Cabibbo-Kobayashi-Maskawa (CKM) matrix\,\footnote{
In case of $\mathcal{C}$, the RH analogue of the quark mixing matrix is equal to the CKM matrix up to overall phases~\cite{Maiezza:2010ic}. }, the ratio  $x_\pi \equiv m_\pi/m_i$, and the decay constant $f_\pi = 130.2\mev$. The kinematic function
\begin{align}
    \beta(a,b,c)\equiv(a^2+b^2+c^2-2ab-2bc-2ca)^{1/2}\;.
\end{align}
The decay rate for a $\nus$ with energy $E^\prime$ is thus $\Gamdec(E^\prime)=\Gamma_{\rm dec,0}/\gamma(E^\prime)$, where $\gamma(E^\prime)=E^\prime/m_{\rm \nus}$ is the boost factor.
The scattering absorption rate $\Gamabs$ is calculated in a manner similar to the production rate as~\cite{Carenza:2023old}
\begin{align}
\Gamabs=\frac{1}{2E^\prime}\left[\int\Pi_n\Pi_e\Pi_p\lvert\mc{M}\rvert^{\prime 2}_{\nus n\to e^- p}f_n(1-f_{e^-})(1-f_p)
+\int\Pi_p\Pi_e\Pi_n\lvert\mc{M}\rvert^{\prime 2}_{\nus p\to e^+ n}f_p(1-f_n)\right]\,.
\end{align}

For completeness, we point out that at the presence of a left-right mixing ($\xi\neq 0$) $\Gamdec$ is modified as $\nus$ acquires additional decay channels. See Appendix~\ref{app:LR-mixing} for more details.

\subsection{The cooling bound}

With the necessary ingredients derived in earlier section, the total energy carried away from the SN by the $\nus$s can be easily derived as
\begin{align}
\label{eq:E_tot_cooling}
E_{\rm cooling}=\int dE_{\nus} dt \int^{\Rcore}_0 4\pi\rprod^2 d\rprod \dfrac{d^2n_{\nus}}{dE_{\nus} dt}P_{\rm srv}(E_{\nus},\rprod;\uu \Renv)\cdot(E_{\nus}\uu\alpha(\rprod)/\alpha(\Renv))\,,
\end{align}
where $\Renv$ is the stellar envelop radius beyond which we consider the $\nus$ as escaped.
We set $\Renv$ as the largest radius in each time slice of the simulation data, typically taking values around $\Renv\sim 10^{13}$ cm.
Note that the expression above contains only a single power of the lapse factor as it calculates the total energy.
If one uses the cooling rate as the criteria for the cooling bound~\cite{raffelt1996stars}, a second lapse factor should be included to take into account the gravitational redshift of time in the differentiation $d/dt$~\cite{Caputo:2022mah}.

\subsection{The energy deposition constraint}

The total energy deposited in the stellar envelop is calculated with the same ingredients as the cooling bound.
For a $\nus$ produced at $\rprod$ with energy $E_{\nus}$, the expectation of its energy deposition at radius $r$ is given by
\begin{align}
P_{\rm srv}(E_{\nus}, \rprod;\uu r)\sum_{i}\Gamma_{i}(E^\prime_{\nus})/\beta_{\nus}(r)\cdot E^\prime_{\nus} f_{{\rm depo},i}\,,
\end{align}
where $i$ runs over the possible deposition channels of $\nus$, $\Gamma_i(E^\prime_{\nus})$ is the decay rate (or scattering absorption rate) of the corresponding channel for a boosted $\nus$ with energy $E^\prime_{\nus}$, and $f_{{\rm depo},i}$ is the fraction of the $\nus$'s energy expected to get deposited.
To calculate $f_{{\rm depo},i}$, we take into account the secondary decays of $\nus$'s daughter particles (\eg, the pions in the hadronic decay channel are subject to further decaying).
The average fractions of mother particle's energy taken away by the daughters are invariant under Lorentz boost.
As stated in the main text, we always consider $e^\pm$ from $\nus$'s primary and secondary decays as the sources of energy deposition, while $\nu/\bar\nu$ are included only when the ambient density is larger than $10^{12}$ g/cm$^3$. For scattering absorption $f_{{\rm depo},i}=1$.

For those $\nus$ gravitationally trapped by the star, we take them to be absorbed at the largest radius they can arrive at (provided the position is beyond $\Rcore$), and determine their absorption mode by comparing $\Gamabs$ and $\Gamdec$ at their maximum radial traveling distance. 
Only when their loss rate $\Gamma$ at the maximum traversing radius satisfying $\Gamma^{-1}>10$ s will the gravitationally trapped $\nus$ be considered as an energy deposition source.
The energy deposited by these trapped $\nus$s are collected denoted as $E_{\rm trapped}$.

The total energy deposited can thus be calculated as
\begin{align}
E_{\rm depo}=&\,\int dE_{\nus} dt \int^{\Rcore}_0 4\pi\rprod^2 d\rprod \dfrac{d^2n}{dE_{\nus} dt}\int^{\Renv}_{\rprod}\dfrac{dl}{\beta(E^\prime_{\nus}(l))}P_{\rm srv}(E_{\nus},\rprod;\uu l)\sum_{i}\Gamma_{i}(E^\prime_{\nus}(l))\cdot E^\prime_{\nus}(l)f_{{\rm depo},i}\nonumber \\
&\,+E_{\rm trapped}\,.\label{eq:E_depo}
\end{align}

\section{Thermal history of sub-GeV RH neutrinos}
\label{app:nus_thermal_history}

In this appendix we briefly discuss the thermal history of the sub-GeV RH neutrino $\nus$ in the mLRSM.
Specifically, we show that $\nus$s are inevitably thermal, and will have a much larger abundance than the observed relic abundance of dark matter.

We first show that freeze-in, where $\nus$ are assumed to be never in equilibrium, is not a viable production mechanism.
The $\nus$ can be produced from the thermal bath via processes like $d+e\to u+\nus$ or $\nu+\nu\to\nus+\nus$, which are mediated by either $W_R$ or $Z_R$.
In the following analysis we ignore the difference between $\mwr$ and $\mzr$ (whose ratio is 1.67 as mentioned earlier), and denote both as $M_V$.
Taking the latter process as an example, the matrix element can be written as
\begin{align}
\lvert\mc{M}\rvert^2=\dfrac{g^4 s^2}{(s-M_V^2)^2+M_V^2\Gamma^2_V}\sim \dfrac{g^4s^2}{M_V\Gamma_{V}}\delta(s-M_V^2)\,,\label{eq:msq_with_pole}
\end{align}
where $s$ is the Mandelstam variable, $g$ is the gauge coupling, $\Gamma_V$ is the decay width, and the narrow-width approximation has been used in the second step. 
The Boltzmann equation thus writes~\cite{Hall:2009bx}
\begin{align}
\dot{n}_{\nus}+3Hn_{\nus}\approx& \dfrac{3T}{128\pi^5}\int ds\dfrac{s}{2\sqrt{s}}\dfrac{(s-\mnus)}{2\sqrt{s}}\dfrac{g^4s^2}{M_V\Gamma_V}\delta(s-M_V^2)\dfrac{K_1(\sqrt{s}/T)}{\sqrt{s}}\\
\approx&\dfrac{3T\uu g^4M_V^4}{512\pi^5\Gamma_V}K_1\left(\dfrac{M_V}{T}\right)\,,
\end{align}
which in terms of the yield $Y=n_{\nus}/s_{\rm plas}$ ($s_{\rm plas}$ is the entropy of the thermal plasma) writes
\begin{align}
\dfrac{dY}{dT}\approx&-\dfrac{3\uu g^4M_V^4}{512\pi^4\, sH\,\Gamma_V}K_1\left(\dfrac{M_V}{T}\right)\,.
\end{align}
The final $\nus$ abundance evaluates to
\begin{align}
Y_\infty\approx& \int^\infty_0 dT \dfrac{dY}{dT}\approx 4\times 10^8\left(\dfrac{g}{0.65}\right)^4\left(\dfrac{g_\ast}{106.75}\right)^{-3/2}\left(\dfrac{1\text{ TeV}}{\Gamma_V}\right)\,,
\end{align}
which is too large a value and suggests the thermalization of $\nus$. 
Similar exercise can be performed for the scenario where the reheating temperature $T_{\rm RH}\ll M_V$ such that the pole in Eq.~\eqref{eq:msq_with_pole} can not be resolved, in which the yield of $\nus$ roughly evaluates to
\begin{align}
Y_\infty\sim\dfrac{M_{\rm pl} T^3_{\rm RH}}{M^4_V}\sim 0.01\left(\dfrac{T_{\rm RH}}{1\text{ GeV}}\right)^3\left(\dfrac{M_V}{100\text{ TeV}}\right)^{-4}\,,
\end{align}
which is still too large compared with the measured dark matter relic abundance.

Meanwhile, the interaction rate between $\nus$ the thermal bath can be estimated as
\begin{align}
\Gamma\sim n\langle\sigma v\rangle\sim N\dfrac{g^4T^5}{\pi^2 M_V^4}\,,\label{eq:thermal_interaction_rate}
\end{align}
where $\langle\sigma v\rangle$ corresponds to either the absorption processes like $\nus+u\leftrightarrow e+d$ or elastic scatterings like $\nus+\nu/l\leftrightarrow\nus+\nu/l$, and $N$ is the number of SM degrees of freedom that can participate in the process.
The temperature where $\nus$ thermally decouple can be estimated by $\Gamma\sim H$, implying
\begin{align}
T_{\rm dec}\sim 5\text{ GeV}\left(\dfrac{M_V}{100\text{ TeV}}\right)^{4/3}\left(\dfrac{g}{0.65}\right)^{-4/3}\left(\dfrac{N}{20}\right)^{-1/3}\,.
\end{align}

The abundance of these thermal $\nus$ at BBN is determined by their freeze-out, where $\nus$ annihilates through $u+\nus\leftrightarrow e+d$ ($n+\nus\to e+p$) or $\nus+\nus\leftrightarrow\nu+\nu$.
The thermal averaged cross section can be roughly estimated as
\begin{align}
\langle\sigma v\rangle\sim N\dfrac{g^4s}{M_V^4}\,,\quad s\sim \text{max}(T^2,m^2_{\nus})\,,
\end{align}
The freeze-out should happen at around $n\langle\sigma v\rangle\sim H$, where $n$ corresponds to the number density of the SM particles for the former process and that of $\nus$ for the latter, taking the values of $T^3/(\pi^2)$ and $m^2_{\nus}TK_2(m_{\nus}/T)/(2\pi^3)$, respectively, with $K_2(z)$ the modified Bessel function of the second kind.
However, it turns out that $n\langle\sigma v\rangle\sim H$ is usually achieved when $T\gg\mnus$, and the number density $n$ for both processes are thus proportional to $T^3$.
With $\langle\sigma v\rangle$ taking a value comparable to that in the interaction rate (cf.Eq.~\eqref{eq:thermal_interaction_rate}), we have the freeze-out temperature $T_{\rm fo}$ to be around $T_{\rm dec}$.
The energy density ratio between $\nus$ and the thermal bath at BBN is thus
\begin{align}
\dfrac{\rho_{\nus}}{\rho_{\rm SM}}=\dfrac{\mnus n_{\nus}}{\frac{\pi^2}{30}g_{\ast,{\rm BBN}}\,T^4_{\rm BBN}}=\dfrac{\mnus \left(\frac{1}{\pi^2}T^3_{\rm fo}\right)\left(\frac{T_{\rm BBN}}{T_{\rm fo}}\right)^3}{\frac{\pi^2}{30}g_{\ast,{\rm BBN}}\,T^4_{\rm BBN}}=\frac{30}{g_{\ast,{\rm BBN}}\pi^4} \dfrac{\mnus}{T_{\rm BBN}}\gg 1\,,
\end{align}
meaning that a non-decaying $\nus$ will also spoil BBN through the influence on the expansion rate.

\section{Results for \tf{$W_L-W_R$}{WL-WR} mixing}
\label{app:LR-mixing}

We now discuss how the model and the corresponding constraints are changed at $\xi\neq 0$. 
Two new decay channels become opened: the photonic decay $\nus\to\nu\gamma$ and the leptonic one $\nus\to l^+l^-\nu$, whose widths can be found in Ref.~\cite{Bondarenko:2018ptm}.
The former arises from the induced neutrino transition dipole moment at one-loop level~\cite{Shrock:1982sc}, and generically dominates the latter.
The hadronic decay width in Eq.~\eqref{eq:Gamma_had} becomes relatively suppressed by a factor of $(1-2\xi/(1+\xi^2))^2$, while is still the the most efficient channel at $\mnus>m_{\pi}$.

\begin{figure}[!tb]
    \centering
    \includegraphics[width=0.5\linewidth]{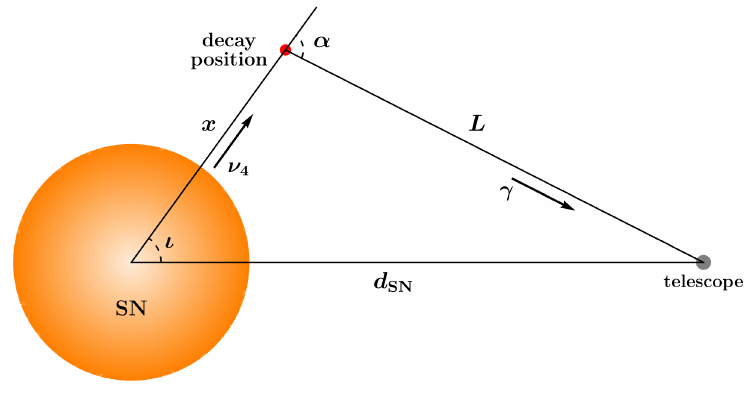}
    \caption{\centering An illustration on the geometry of $\nus$'s photonic decay.}
    \label{fig:decay_geo}
\end{figure}

Then regarding the SNe, $\nus$'s photonic decay may generate a $\gamma$-ray afterglow following the neutrinos' arrival, a new signal absent in Sec.~\ref{sec:sn-constraint}.
The derivation of the $\gamma$-ray constraint is less direct compared with the cooling bound and the energy deposition bound due to its geometry.
Let the direction from the SN to the telescope be the polar axis, $\iota$ the azimuthal angle of the escaped $\nus$'s propagation direction, $\alpha$ the angle between the $\nus$'s propagation direction and it's daughter photon's. And denote $d_{\rm SN}$, $x$ and $L$ as the distances from the telescope to the SN, the $\nus$'s traveling distance before its decay, and the distance between the telescope and the decay position of the $\nus$, respectively.
The aforementioned quantities and the corresponding geometry is illustrated in Fig.~\ref{fig:decay_geo}.
The time delay between the neutrino burst and the photon from the $\nus$'s decay is then
\begin{align}
\delta t=x/\beta+L-d_{\rm NS}\,,
\end{align}
where $\beta$ is the velocity of the $\nus$.
Since the parameter space of our interest always implies $x\ll d_{\rm NS}$, the angles $\iota$ and $\alpha$ are almost identical, and the time delay can be expressed as
\begin{align}\label{eq:delta_t_approx}
\delta t\approx x(1/\beta-\cos\alpha)\approx x(1/\beta-\cos\iota)\,.
\end{align}
The orientation of the daughter photon is isotropic in the rest frame of the $\nus$, and hence the probability that in the observer frame the daughter photon is oriented right at the telescope is
\begin{align}
P_\gamma(\iota, x)=\dfrac{1}{4\pi\gamma^2(1-\beta\cos\alpha)^2}\,,
\end{align}
where $\gamma$ is the Lorentz factor of the $\nus$.
The energy of the daughter photon $E_\gamma$ is expressed with the geometric parameters and the Lorentz factors as
\begin{align}\label{eq:Egamma_as_alpha}
E_\gamma&=\dfrac{\mnus}{2\gamma(1+\beta\cos\alpha)}\,.
\end{align}
With the lower and upper bound of the telescope's energy band denoted as $E_{\gamma,{\rm min}}$ and $E_{\gamma,{\rm max}}$, respectively, and the observation time of the telescope $T_{\rm obs}$, the total photon number observed by the telescope is therefore given by
\begin{align}
\begin{aligned}
N_{\gamma}=\int& \sin\iota\uu d\iota \int^{\Rcore}_0 2\pi r^2 dr \int dE_{\nus}dt \dfrac{d^2n}{dE_{\nus}dt} P_{\rm srv}(E_{\nus},r;\; \Renv) \int^\infty_0 \underbrace{e^{-\Gamdec x/\beta}}_{\text{remnant flux fraction}}\underbrace{\dfrac{dx}{\beta}\Gamdec{\rm Br}_{\rm ph.}}_{\text{photonic decay probability}}\\
&\times P_\gamma(\iota, x)\times \underbrace{\dfrac{A_{\rm tele}}{4\pi L^2}}_{\text{telescope's solid angle}}\times\Theta(T_{\rm obs}-\delta t)\Theta(E_{\gamma,{\rm max}}-E_\gamma)\Theta(E_\gamma-E_{\gamma,{\rm min}})\,,
\end{aligned}
\end{align}
where $\Theta(z)$ is the Heaviside function.
Since the angles $\iota$ and $\alpha$ are almost identical, as stated earlier, the $\iota$ integration as well as the $\alpha$-dependence in $P_\gamma$ can be rewritten with $E_\gamma$ using Eq.~\eqref{eq:Egamma_as_alpha}, leading to
\begin{align}
\begin{aligned}
N_\gamma=&\dfrac{A_{\rm tele}}{16\pi^2\pi d^2_{SN}}\int dE_{\nus} dt \int^{\Rcore}_0 4\pi r^2 dr \dfrac{d^2n}{dE_{\nus} dt}P_{\rm srv}(E_{\nus},r;\; \Renv)\\
&\cdot \int^{E_{\rm max}}_{E_{\rm min}}\dfrac{dE_\gamma}{\beta\uu\gamma\uu \mnus}\int^{x_{\rm max}}_0 \dfrac{dx}{\beta}\uu e^{-\Gamdec x/\beta} \uu \Gamdec \uu {\rm Br_{\rm ph.}}\,,
\end{aligned}\label{eq:Ngamma_simplified}
\end{align}
where $x_{\rm max}=2\beta\gamma E_\gamma T_{\rm obs}/\mnus$ is given by the $\delta t$ upper bound using Eq.~\eqref{eq:delta_t_approx}, $E_{\rm max}={\rm min}\left[100\uu{\rm MeV},\gamma(1+\beta)(m_\nu/2)\right]$ and $E_{\rm min}={\rm max}\left[25\uu{\rm MeV},\gamma(1-\beta)(m_\nu/2)\right]$ are the maximal and minimal energy of the photon required by both the telescope and the initial $\nus$.
The leading order approximation $L\approx d_{\rm SN}$ has also been taken for the solid angle of the telescope.
The integration on the second line of Eq.~\eqref{eq:Ngamma_simplified} can be performed analytically, which evaluates to
\begin{align}\label{eq:Ex_integration}
\int^{E_{\rm max}}_{E_{\rm min}}\dfrac{dE_\gamma}{\beta\uu\gamma\uu \mnus}\int^{x_{\rm max}}_0 \dfrac{dx}{\beta}\uu e^{-\Gamdec x/\beta} \uu \Gamdec \uu {\rm Br_{\rm ph.}}=\dfrac{E_{\rm max}-E_{\rm min}-E_0\left(e^{-E_{\rm min}/E_0}-e^{-E_{\rm max}/E_0}\right)}{\beta\uu\gamma\uu \mnus} {\rm Br_{\rm ph.}}\,,
\end{align}
where $E_0=\mnus/(2\gamma\uu T_{\rm obs}\Gamdec)$.

The quantity $\Phi_\gamma=N_\gamma/A_{\rm tele}$ is usually used for the $\gamma$-ray constraints. 
Specifically for SN1987A, the Solar Maximum Mission's (SMM) measurement of 25-100 MeV photons since the neutrino burst constraints the $\gamma$-ray flux from $\nus$ to be $\Phi_\gamma\lesssim 1.38\,\text{cm}^{-2}$~\cite{Chauhan:2025mnn,Brdar:2023tmi} at the $2\sigma$-level.
This constraint may be tightened with a current or future telescope of better sensitivity~\cite{Fermilat,e-ASTROGAM:2017pxr,Moiseev:2015lva,Wu:2014tya} observing an SN in the Galaxy.
For example, assuming 3600s of observation time, Fermi-LAT is expected to achieve $\Phi_\gamma\lesssim 5.2\times 10^{-4}{\rm cm}^{-2}$~\cite{Jaeckel:2017tud} for photons more energetic than 100 MeV.
However, this new signal is still generically overshadowed by the cosmological ones.

For existing constraints, the SN energy deposition constraint may limit a slightly smaller range of $\mnus$ due to the suppression of the hadronic decay width.
For the cosmological constraints, at $\mnus>m_{\pi}$ the $\tau_{\nus}$ bound now corresponds to a slightly lower $\mwr$ due the hadronic decay width's suppression.
At $\mnus<m_{\pi}$, again the existing constraints do not directly apply due to the decay mode discrepancy.
The most relevant constraint, as far as we know, comes from the magnetic moment portal neutrino~\cite{Brdar:2020quo,Brdar:2023tmi} where the $\nus$s can decay photonically through $\frac{\mu^\alpha_\nu}{2}F_{\mu\nu}\bar{\nu}^\alpha_L\sigma^{\mu\nu}\nu_R$, which in this mass range requires the magnetic moment $\mu_\nu\lesssim 10^{-12}$-$10^{-13}e/(2m_e)$~\cite{Brdar:2020quo}.
After combining with the $\nldb$ decay constraints which are stronger for $\xi\neq 0$ (due to long-range pion exchange~\cite{Prezeau:2003xn,deVries:2022nyh}), no parameter space remains unexplored in this $\mnus$ range after the operation of future $\nldb$ experiments.
Our conclusions for $\xi=0$ thus remains unchanged.

\bibliographystyle{apsrev4-1}

\bibliography{reference}

\begin{thebibliography}{114}%
\makeatletter
\providecommand \@ifxundefined [1]{%
 \@ifx{#1\undefined}
}%
\providecommand \@ifnum [1]{%
 \ifnum #1\expandafter \@firstoftwo
 \else \expandafter \@secondoftwo
 \fi
}%
\providecommand \@ifx [1]{%
 \ifx #1\expandafter \@firstoftwo
 \else \expandafter \@secondoftwo
 \fi
}%
\providecommand \natexlab [1]{#1}%
\providecommand \enquote  [1]{``#1''}%
\providecommand \bibnamefont  [1]{#1}%
\providecommand \bibfnamefont [1]{#1}%
\providecommand \citenamefont [1]{#1}%
\providecommand \href@noop [0]{\@secondoftwo}%
\providecommand \href [0]{\begingroup \@sanitize@url \@href}%
\providecommand \@href[1]{\@@startlink{#1}\@@href}%
\providecommand \@@href[1]{\endgroup#1\@@endlink}%
\providecommand \@sanitize@url [0]{\catcode `\\12\catcode `\$12\catcode
  `\&12\catcode `\#12\catcode `\^12\catcode `\_12\catcode `\%12\relax}%
\providecommand \@@startlink[1]{}%
\providecommand \@@endlink[0]{}%
\providecommand \url  [0]{\begingroup\@sanitize@url \@url }%
\providecommand \@url [1]{\endgroup\@href {#1}{\urlprefix }}%
\providecommand \urlprefix  [0]{URL }%
\providecommand \Eprint [0]{\href }%
\providecommand \doibase [0]{http://dx.doi.org/}%
\providecommand \selectlanguage [0]{\@gobble}%
\providecommand \bibinfo  [0]{\@secondoftwo}%
\providecommand \bibfield  [0]{\@secondoftwo}%
\providecommand \translation [1]{[#1]}%
\providecommand \BibitemOpen [0]{}%
\providecommand \bibitemStop [0]{}%
\providecommand \bibitemNoStop [0]{.\EOS\space}%
\providecommand \EOS [0]{\spacefactor3000\relax}%
\providecommand \BibitemShut  [1]{\csname bibitem#1\endcsname}%
\let\auto@bib@innerbib\@empty
\bibitem [{\citenamefont {Minkowski}(1977)}]{Minkowski:1977sc}%
  \BibitemOpen
  \bibfield  {author} {\bibinfo {author} {\bibfnamefont {P.}~\bibnamefont
  {Minkowski}},\ }\href {\doibase 10.1016/0370-2693(77)90435-X} {\bibfield
  {journal} {\bibinfo  {journal} {Phys. Lett. B}\ }\textbf {\bibinfo {volume}
  {67}},\ \bibinfo {pages} {421} (\bibinfo {year} {1977})}\BibitemShut
  {NoStop}%
\bibitem [{\citenamefont {Yanagida}(1979)}]{Yanagida:1979as}%
  \BibitemOpen
  \bibfield  {author} {\bibinfo {author} {\bibfnamefont {T.}~\bibnamefont
  {Yanagida}},\ }\href@noop {} {\bibfield  {journal} {\bibinfo  {journal}
  {Conf. Proc. C}\ }\textbf {\bibinfo {volume} {7902131}},\ \bibinfo {pages}
  {95} (\bibinfo {year} {1979})}\BibitemShut {NoStop}%
\bibitem [{\citenamefont {Gell-Mann}\ \emph {et~al.}(1979)\citenamefont
  {Gell-Mann}, \citenamefont {Ramond},\ and\ \citenamefont
  {Slansky}}]{Gell-Mann:1979vob}%
  \BibitemOpen
  \bibfield  {author} {\bibinfo {author} {\bibfnamefont {M.}~\bibnamefont
  {Gell-Mann}}, \bibinfo {author} {\bibfnamefont {P.}~\bibnamefont {Ramond}}, \
  and\ \bibinfo {author} {\bibfnamefont {R.}~\bibnamefont {Slansky}},\
  }\href@noop {} {\bibfield  {journal} {\bibinfo  {journal} {Conf. Proc. C}\
  }\textbf {\bibinfo {volume} {790927}},\ \bibinfo {pages} {315} (\bibinfo
  {year} {1979})},\ \Eprint {http://arxiv.org/abs/1306.4669} {arXiv:1306.4669
  [hep-th]} \BibitemShut {NoStop}%
\bibitem [{\citenamefont {Mohapatra}\ and\ \citenamefont
  {Senjanovic}(1980)}]{Mohapatra:1979ia}%
  \BibitemOpen
  \bibfield  {author} {\bibinfo {author} {\bibfnamefont {R.~N.}\ \bibnamefont
  {Mohapatra}}\ and\ \bibinfo {author} {\bibfnamefont {G.}~\bibnamefont
  {Senjanovic}},\ }\href {\doibase 10.1103/PhysRevLett.44.912} {\bibfield
  {journal} {\bibinfo  {journal} {Phys. Rev. Lett.}\ }\textbf {\bibinfo
  {volume} {44}},\ \bibinfo {pages} {912} (\bibinfo {year} {1980})}\BibitemShut
  {NoStop}%
\bibitem [{\citenamefont {Glashow}(1980)}]{Glashow:1979nm}%
  \BibitemOpen
  \bibfield  {author} {\bibinfo {author} {\bibfnamefont {S.~L.}\ \bibnamefont
  {Glashow}},\ }\href {\doibase 10.1007/978-1-4684-7197-7_15} {\bibfield
  {journal} {\bibinfo  {journal} {NATO Sci. Ser. B}\ }\textbf {\bibinfo
  {volume} {61}},\ \bibinfo {pages} {687} (\bibinfo {year} {1980})}\BibitemShut
  {NoStop}%
\bibitem [{\citenamefont {Pati}\ and\ \citenamefont
  {Salam}(1974)}]{Pati:1974yy}%
  \BibitemOpen
  \bibfield  {author} {\bibinfo {author} {\bibfnamefont {J.~C.}\ \bibnamefont
  {Pati}}\ and\ \bibinfo {author} {\bibfnamefont {A.}~\bibnamefont {Salam}},\
  }\href {\doibase 10.1103/PhysRevD.10.275} {\bibfield  {journal} {\bibinfo
  {journal} {Phys. Rev. D}\ }\textbf {\bibinfo {volume} {10}},\ \bibinfo
  {pages} {275} (\bibinfo {year} {1974})},\ \bibinfo {note} {[Erratum:
  Phys.Rev.D 11, 703--703 (1975)]}\BibitemShut {NoStop}%
\bibitem [{\citenamefont {Mohapatra}\ and\ \citenamefont
  {Pati}(1975{\natexlab{a}})}]{Mohapatra:1974gc}%
  \BibitemOpen
  \bibfield  {author} {\bibinfo {author} {\bibfnamefont {R.~N.}\ \bibnamefont
  {Mohapatra}}\ and\ \bibinfo {author} {\bibfnamefont {J.~C.}\ \bibnamefont
  {Pati}},\ }\href {\doibase 10.1103/PhysRevD.11.2558} {\bibfield  {journal}
  {\bibinfo  {journal} {Phys. Rev. D}\ }\textbf {\bibinfo {volume} {11}},\
  \bibinfo {pages} {2558} (\bibinfo {year} {1975}{\natexlab{a}})}\BibitemShut
  {NoStop}%
\bibitem [{\citenamefont {Mohapatra}\ and\ \citenamefont
  {Pati}(1975{\natexlab{b}})}]{Mohapatra:1974hk}%
  \BibitemOpen
  \bibfield  {author} {\bibinfo {author} {\bibfnamefont {R.~N.}\ \bibnamefont
  {Mohapatra}}\ and\ \bibinfo {author} {\bibfnamefont {J.~C.}\ \bibnamefont
  {Pati}},\ }\href {\doibase 10.1103/PhysRevD.11.566} {\bibfield  {journal}
  {\bibinfo  {journal} {Phys. Rev. D}\ }\textbf {\bibinfo {volume} {11}},\
  \bibinfo {pages} {566} (\bibinfo {year} {1975}{\natexlab{b}})}\BibitemShut
  {NoStop}%
\bibitem [{\citenamefont {Senjanovic}\ and\ \citenamefont
  {Mohapatra}(1975)}]{Senjanovic:1975rk}%
  \BibitemOpen
  \bibfield  {author} {\bibinfo {author} {\bibfnamefont {G.}~\bibnamefont
  {Senjanovic}}\ and\ \bibinfo {author} {\bibfnamefont {R.~N.}\ \bibnamefont
  {Mohapatra}},\ }\href {\doibase 10.1103/PhysRevD.12.1502} {\bibfield
  {journal} {\bibinfo  {journal} {Phys. Rev. D}\ }\textbf {\bibinfo {volume}
  {12}},\ \bibinfo {pages} {1502} (\bibinfo {year} {1975})}\BibitemShut
  {NoStop}%
\bibitem [{\citenamefont {Senjanovic}(1979)}]{Senjanovic:1978ev}%
  \BibitemOpen
  \bibfield  {author} {\bibinfo {author} {\bibfnamefont {G.}~\bibnamefont
  {Senjanovic}},\ }\href {\doibase 10.1016/0550-3213(79)90604-7} {\bibfield
  {journal} {\bibinfo  {journal} {Nucl. Phys. B}\ }\textbf {\bibinfo {volume}
  {153}},\ \bibinfo {pages} {334} (\bibinfo {year} {1979})}\BibitemShut
  {NoStop}%
\bibitem [{\citenamefont {Mohapatra}\ and\ \citenamefont
  {Senjanovic}(1981)}]{Mohapatra:1980yp}%
  \BibitemOpen
  \bibfield  {author} {\bibinfo {author} {\bibfnamefont {R.~N.}\ \bibnamefont
  {Mohapatra}}\ and\ \bibinfo {author} {\bibfnamefont {G.}~\bibnamefont
  {Senjanovic}},\ }\href {\doibase 10.1103/PhysRevD.23.165} {\bibfield
  {journal} {\bibinfo  {journal} {Phys. Rev. D}\ }\textbf {\bibinfo {volume}
  {23}},\ \bibinfo {pages} {165} (\bibinfo {year} {1981})}\BibitemShut
  {NoStop}%
\bibitem [{\citenamefont {Senjanovic}(2011)}]{Senjanovic:2011zz}%
  \BibitemOpen
  \bibfield  {author} {\bibinfo {author} {\bibfnamefont {G.}~\bibnamefont
  {Senjanovic}},\ }\href {\doibase 10.1393/ncr/i2011-10061-8} {\bibfield
  {journal} {\bibinfo  {journal} {Riv. Nuovo Cim.}\ }\textbf {\bibinfo {volume}
  {34}},\ \bibinfo {pages} {1} (\bibinfo {year} {2011})}\BibitemShut {NoStop}%
\bibitem [{\citenamefont {Keung}\ and\ \citenamefont
  {Senjanovic}(1983)}]{Keung:1983uu}%
  \BibitemOpen
  \bibfield  {author} {\bibinfo {author} {\bibfnamefont {W.-Y.}\ \bibnamefont
  {Keung}}\ and\ \bibinfo {author} {\bibfnamefont {G.}~\bibnamefont
  {Senjanovic}},\ }\href {\doibase 10.1103/PhysRevLett.50.1427} {\bibfield
  {journal} {\bibinfo  {journal} {Phys. Rev. Lett.}\ }\textbf {\bibinfo
  {volume} {50}},\ \bibinfo {pages} {1427} (\bibinfo {year}
  {1983})}\BibitemShut {NoStop}%
\bibitem [{\citenamefont {Aad}\ \emph {et~al.}(2023)\citenamefont {Aad} \emph
  {et~al.}}]{ATLAS:2023cjo}%
  \BibitemOpen
  \bibfield  {author} {\bibinfo {author} {\bibfnamefont {G.}~\bibnamefont
  {Aad}} \emph {et~al.} (\bibinfo {collaboration} {ATLAS}),\ }\href {\doibase
  10.1140/epjc/s10052-023-12021-9} {\bibfield  {journal} {\bibinfo  {journal}
  {Eur. Phys. J. C}\ }\textbf {\bibinfo {volume} {83}},\ \bibinfo {pages}
  {1164} (\bibinfo {year} {2023})},\ \Eprint {http://arxiv.org/abs/2304.09553}
  {arXiv:2304.09553 [hep-ex]} \BibitemShut {NoStop}%
\bibitem [{\citenamefont {Aad}\ \emph {et~al.}(2019{\natexlab{a}})\citenamefont
  {Aad} \emph {et~al.}}]{ATLAS:2019lsy}%
  \BibitemOpen
  \bibfield  {author} {\bibinfo {author} {\bibfnamefont {G.}~\bibnamefont
  {Aad}} \emph {et~al.} (\bibinfo {collaboration} {ATLAS}),\ }\href {\doibase
  10.1103/PhysRevD.100.052013} {\bibfield  {journal} {\bibinfo  {journal}
  {Phys. Rev. D}\ }\textbf {\bibinfo {volume} {100}},\ \bibinfo {pages}
  {052013} (\bibinfo {year} {2019}{\natexlab{a}})},\ \Eprint
  {http://arxiv.org/abs/1906.05609} {arXiv:1906.05609 [hep-ex]} \BibitemShut
  {NoStop}%
\bibitem [{\citenamefont {Tumasyan}\ \emph {et~al.}(2022)\citenamefont
  {Tumasyan} \emph {et~al.}}]{CMS:2022krd}%
  \BibitemOpen
  \bibfield  {author} {\bibinfo {author} {\bibfnamefont {A.}~\bibnamefont
  {Tumasyan}} \emph {et~al.} (\bibinfo {collaboration} {CMS}),\ }\href
  {\doibase 10.1007/JHEP07(2022)067} {\bibfield  {journal} {\bibinfo  {journal}
  {JHEP}\ }\textbf {\bibinfo {volume} {07}},\ \bibinfo {pages} {067} (\bibinfo
  {year} {2022})},\ \Eprint {http://arxiv.org/abs/2202.06075} {arXiv:2202.06075
  [hep-ex]} \BibitemShut {NoStop}%
\bibitem [{\citenamefont {de~Vries}\ \emph {et~al.}(2025)\citenamefont
  {de~Vries}, \citenamefont {Dreiner}, \citenamefont {Groot}, \citenamefont
  {G{\"u}nther},\ and\ \citenamefont {Wang}}]{deVries:2024mla}%
  \BibitemOpen
  \bibfield  {author} {\bibinfo {author} {\bibfnamefont {J.}~\bibnamefont
  {de~Vries}}, \bibinfo {author} {\bibfnamefont {H.~K.}\ \bibnamefont
  {Dreiner}}, \bibinfo {author} {\bibfnamefont {J.}~\bibnamefont {Groot}},
  \bibinfo {author} {\bibfnamefont {J.~Y.}\ \bibnamefont {G{\"u}nther}}, \ and\
  \bibinfo {author} {\bibfnamefont {Z.~S.}\ \bibnamefont {Wang}},\ }\href
  {\doibase 10.1007/JHEP04(2025)007} {\bibfield  {journal} {\bibinfo  {journal}
  {JHEP}\ }\textbf {\bibinfo {volume} {04}},\ \bibinfo {pages} {007} (\bibinfo
  {year} {2025})},\ \Eprint {http://arxiv.org/abs/2406.15091} {arXiv:2406.15091
  [hep-ph]} \BibitemShut {NoStop}%
\bibitem [{\citenamefont {Nemev{\v{s}}ek}\ \emph {et~al.}(2018)\citenamefont
  {Nemev{\v{s}}ek}, \citenamefont {Nesti},\ and\ \citenamefont
  {Popara}}]{Nemevsek:2018bbt}%
  \BibitemOpen
  \bibfield  {author} {\bibinfo {author} {\bibfnamefont {M.}~\bibnamefont
  {Nemev{\v{s}}ek}}, \bibinfo {author} {\bibfnamefont {F.}~\bibnamefont
  {Nesti}}, \ and\ \bibinfo {author} {\bibfnamefont {G.}~\bibnamefont
  {Popara}},\ }\href {\doibase 10.1103/PhysRevD.97.115018} {\bibfield
  {journal} {\bibinfo  {journal} {Phys. Rev. D}\ }\textbf {\bibinfo {volume}
  {97}},\ \bibinfo {pages} {115018} (\bibinfo {year} {2018})},\ \Eprint
  {http://arxiv.org/abs/1801.05813} {arXiv:1801.05813 [hep-ph]} \BibitemShut
  {NoStop}%
\bibitem [{\citenamefont {Helo}\ \emph {et~al.}(2014)\citenamefont {Helo},
  \citenamefont {Hirsch},\ and\ \citenamefont {Kovalenko}}]{Helo:2013esa}%
  \BibitemOpen
  \bibfield  {author} {\bibinfo {author} {\bibfnamefont {J.~C.}\ \bibnamefont
  {Helo}}, \bibinfo {author} {\bibfnamefont {M.}~\bibnamefont {Hirsch}}, \ and\
  \bibinfo {author} {\bibfnamefont {S.}~\bibnamefont {Kovalenko}},\ }\href
  {\doibase 10.1103/PhysRevD.89.073005} {\bibfield  {journal} {\bibinfo
  {journal} {Phys. Rev. D}\ }\textbf {\bibinfo {volume} {89}},\ \bibinfo
  {pages} {073005} (\bibinfo {year} {2014})},\ \bibinfo {note} {[Erratum:
  Phys.Rev.D 93, 099902 (2016)]},\ \Eprint {http://arxiv.org/abs/1312.2900}
  {arXiv:1312.2900 [hep-ph]} \BibitemShut {NoStop}%
\bibitem [{\citenamefont {Abgrall}\ \emph {et~al.}(2021)\citenamefont {Abgrall}
  \emph {et~al.}}]{LEGEND:2021bnm}%
  \BibitemOpen
  \bibfield  {author} {\bibinfo {author} {\bibfnamefont {N.}~\bibnamefont
  {Abgrall}} \emph {et~al.} (\bibinfo {collaboration} {LEGEND}),\ }\href@noop
  {} {\  (\bibinfo {year} {2021})},\ \Eprint {http://arxiv.org/abs/2107.11462}
  {arXiv:2107.11462 [physics.ins-det]} \BibitemShut {NoStop}%
\bibitem [{\citenamefont {Adhikari}\ \emph {et~al.}(2022)\citenamefont
  {Adhikari} \emph {et~al.}}]{nEXO:2021ujk}%
  \BibitemOpen
  \bibfield  {author} {\bibinfo {author} {\bibfnamefont {G.}~\bibnamefont
  {Adhikari}} \emph {et~al.} (\bibinfo {collaboration} {nEXO}),\ }\href
  {\doibase 10.1088/1361-6471/ac3631} {\bibfield  {journal} {\bibinfo
  {journal} {J. Phys. G}\ }\textbf {\bibinfo {volume} {49}},\ \bibinfo {pages}
  {015104} (\bibinfo {year} {2022})},\ \Eprint
  {http://arxiv.org/abs/2106.16243} {arXiv:2106.16243 [nucl-ex]} \BibitemShut
  {NoStop}%
\bibitem [{\citenamefont {Agostini}\ \emph {et~al.}(2023)\citenamefont
  {Agostini}, \citenamefont {Benato}, \citenamefont {Detwiler}, \citenamefont
  {Men{\'e}ndez},\ and\ \citenamefont {Vissani}}]{Agostini:2022zub}%
  \BibitemOpen
  \bibfield  {author} {\bibinfo {author} {\bibfnamefont {M.}~\bibnamefont
  {Agostini}}, \bibinfo {author} {\bibfnamefont {G.}~\bibnamefont {Benato}},
  \bibinfo {author} {\bibfnamefont {J.~A.}\ \bibnamefont {Detwiler}}, \bibinfo
  {author} {\bibfnamefont {J.}~\bibnamefont {Men{\'e}ndez}}, \ and\ \bibinfo
  {author} {\bibfnamefont {F.}~\bibnamefont {Vissani}},\ }\href {\doibase
  10.1103/RevModPhys.95.025002} {\bibfield  {journal} {\bibinfo  {journal}
  {Rev. Mod. Phys.}\ }\textbf {\bibinfo {volume} {95}},\ \bibinfo {pages}
  {025002} (\bibinfo {year} {2023})},\ \Eprint
  {http://arxiv.org/abs/2202.01787} {arXiv:2202.01787 [hep-ex]} \BibitemShut
  {NoStop}%
\bibitem [{\citenamefont {de~Vries}\ \emph {et~al.}(2022)\citenamefont
  {de~Vries}, \citenamefont {Li}, \citenamefont {Ramsey-Musolf},\ and\
  \citenamefont {Vasquez}}]{deVries:2022nyh}%
  \BibitemOpen
  \bibfield  {author} {\bibinfo {author} {\bibfnamefont {J.}~\bibnamefont
  {de~Vries}}, \bibinfo {author} {\bibfnamefont {G.}~\bibnamefont {Li}},
  \bibinfo {author} {\bibfnamefont {M.~J.}\ \bibnamefont {Ramsey-Musolf}}, \
  and\ \bibinfo {author} {\bibfnamefont {J.~C.}\ \bibnamefont {Vasquez}},\
  }\href {\doibase 10.1007/JHEP11(2022)056} {\bibfield  {journal} {\bibinfo
  {journal} {JHEP}\ }\textbf {\bibinfo {volume} {11}},\ \bibinfo {pages} {056}
  (\bibinfo {year} {2022})},\ \Eprint {http://arxiv.org/abs/2209.03031}
  {arXiv:2209.03031 [hep-ph]} \BibitemShut {NoStop}%
\bibitem [{\citenamefont {Barbieri}\ and\ \citenamefont
  {Mohapatra}(1989)}]{Barbieri:1988av}%
  \BibitemOpen
  \bibfield  {author} {\bibinfo {author} {\bibfnamefont {R.}~\bibnamefont
  {Barbieri}}\ and\ \bibinfo {author} {\bibfnamefont {R.~N.}\ \bibnamefont
  {Mohapatra}},\ }\href {\doibase 10.1103/PhysRevD.39.1229} {\bibfield
  {journal} {\bibinfo  {journal} {Phys. Rev. D}\ }\textbf {\bibinfo {volume}
  {39}},\ \bibinfo {pages} {1229} (\bibinfo {year} {1989})}\BibitemShut
  {NoStop}%
\bibitem [{\citenamefont {Alves}\ \emph {et~al.}(2024)\citenamefont {Alves},
  \citenamefont {Fong}, \citenamefont {Leal},\ and\ \citenamefont
  {Zukanovich~Funchal}}]{Alves:2023znq}%
  \BibitemOpen
  \bibfield  {author} {\bibinfo {author} {\bibfnamefont {G.~F.~S.}\
  \bibnamefont {Alves}}, \bibinfo {author} {\bibfnamefont {C.~S.}\ \bibnamefont
  {Fong}}, \bibinfo {author} {\bibfnamefont {L.~P.~S.}\ \bibnamefont {Leal}}, \
  and\ \bibinfo {author} {\bibfnamefont {R.}~\bibnamefont
  {Zukanovich~Funchal}},\ }\href {\doibase 10.1103/PhysRevLett.133.161802}
  {\bibfield  {journal} {\bibinfo  {journal} {Phys. Rev. Lett.}\ }\textbf
  {\bibinfo {volume} {133}},\ \bibinfo {pages} {161802} (\bibinfo {year}
  {2024})},\ \Eprint {http://arxiv.org/abs/2307.04862} {arXiv:2307.04862
  [hep-ph]} \BibitemShut {NoStop}%
\bibitem [{\citenamefont {Group}(2025)}]{pdg_WR}%
  \BibitemOpen
  \bibfield  {author} {\bibinfo {author} {\bibfnamefont {P.~D.}\ \bibnamefont
  {Group}},\ }\href@noop {} {\enquote {\bibinfo {title} {{${{\mathit
  W}_{{{R}}}}$ (Right-Handed ${{\mathit W}}$ Boson) MASS LIMITS}},}\ }\bibinfo
  {howpublished} {\url{https://pdglive.lbl.gov//DataBlock.action?node=S056MR}}
  (\bibinfo {year} {2025})\BibitemShut {NoStop}%
\bibitem [{\citenamefont {Abe}\ \emph {et~al.}(2023)\citenamefont {Abe} \emph
  {et~al.}}]{KamLAND-Zen:2022tow}%
  \BibitemOpen
  \bibfield  {author} {\bibinfo {author} {\bibfnamefont {S.}~\bibnamefont
  {Abe}} \emph {et~al.} (\bibinfo {collaboration} {KamLAND-Zen}),\ }\href
  {\doibase 10.1103/PhysRevLett.130.051801} {\bibfield  {journal} {\bibinfo
  {journal} {Phys. Rev. Lett.}\ }\textbf {\bibinfo {volume} {130}},\ \bibinfo
  {pages} {051801} (\bibinfo {year} {2023})},\ \Eprint
  {http://arxiv.org/abs/2203.02139} {arXiv:2203.02139 [hep-ex]} \BibitemShut
  {NoStop}%
\bibitem [{\citenamefont {Abe}\ \emph {et~al.}(2024)\citenamefont {Abe} \emph
  {et~al.}}]{KamLAND-Zen:2024eml}%
  \BibitemOpen
  \bibfield  {author} {\bibinfo {author} {\bibfnamefont {S.}~\bibnamefont
  {Abe}} \emph {et~al.} (\bibinfo {collaboration} {KamLAND-Zen}),\ }\href@noop
  {} {\  (\bibinfo {year} {2024})},\ \Eprint {http://arxiv.org/abs/2406.11438}
  {arXiv:2406.11438 [hep-ex]} \BibitemShut {NoStop}%
\bibitem [{\citenamefont {Maiezza}\ \emph {et~al.}(2010)\citenamefont
  {Maiezza}, \citenamefont {Nemevsek}, \citenamefont {Nesti},\ and\
  \citenamefont {Senjanovic}}]{Maiezza:2010ic}%
  \BibitemOpen
  \bibfield  {author} {\bibinfo {author} {\bibfnamefont {A.}~\bibnamefont
  {Maiezza}}, \bibinfo {author} {\bibfnamefont {M.}~\bibnamefont {Nemevsek}},
  \bibinfo {author} {\bibfnamefont {F.}~\bibnamefont {Nesti}}, \ and\ \bibinfo
  {author} {\bibfnamefont {G.}~\bibnamefont {Senjanovic}},\ }\href {\doibase
  10.1103/PhysRevD.82.055022} {\bibfield  {journal} {\bibinfo  {journal} {Phys.
  Rev. D}\ }\textbf {\bibinfo {volume} {82}},\ \bibinfo {pages} {055022}
  (\bibinfo {year} {2010})},\ \Eprint {http://arxiv.org/abs/1005.5160}
  {arXiv:1005.5160 [hep-ph]} \BibitemShut {NoStop}%
\bibitem [{\citenamefont {Chang}\ \emph {et~al.}(1984)\citenamefont {Chang},
  \citenamefont {Mohapatra},\ and\ \citenamefont {Parida}}]{Chang:1983fu}%
  \BibitemOpen
  \bibfield  {author} {\bibinfo {author} {\bibfnamefont {D.}~\bibnamefont
  {Chang}}, \bibinfo {author} {\bibfnamefont {R.~N.}\ \bibnamefont
  {Mohapatra}}, \ and\ \bibinfo {author} {\bibfnamefont {M.~K.}\ \bibnamefont
  {Parida}},\ }\href {\doibase 10.1103/PhysRevLett.52.1072} {\bibfield
  {journal} {\bibinfo  {journal} {Phys. Rev. Lett.}\ }\textbf {\bibinfo
  {volume} {52}},\ \bibinfo {pages} {1072} (\bibinfo {year}
  {1984})}\BibitemShut {NoStop}%
\bibitem [{\citenamefont {Nemevsek}\ \emph {et~al.}(2013)\citenamefont
  {Nemevsek}, \citenamefont {Senjanovic},\ and\ \citenamefont
  {Tello}}]{Nemevsek:2012iq}%
  \BibitemOpen
  \bibfield  {author} {\bibinfo {author} {\bibfnamefont {M.}~\bibnamefont
  {Nemevsek}}, \bibinfo {author} {\bibfnamefont {G.}~\bibnamefont
  {Senjanovic}}, \ and\ \bibinfo {author} {\bibfnamefont {V.}~\bibnamefont
  {Tello}},\ }\href {\doibase 10.1103/PhysRevLett.110.151802} {\bibfield
  {journal} {\bibinfo  {journal} {Phys. Rev. Lett.}\ }\textbf {\bibinfo
  {volume} {110}},\ \bibinfo {pages} {151802} (\bibinfo {year} {2013})},\
  \Eprint {http://arxiv.org/abs/1211.2837} {arXiv:1211.2837 [hep-ph]}
  \BibitemShut {NoStop}%
\bibitem [{\citenamefont {Bertolini}\ \emph {et~al.}(2014)\citenamefont
  {Bertolini}, \citenamefont {Maiezza},\ and\ \citenamefont
  {Nesti}}]{Bertolini:2014sua}%
  \BibitemOpen
  \bibfield  {author} {\bibinfo {author} {\bibfnamefont {S.}~\bibnamefont
  {Bertolini}}, \bibinfo {author} {\bibfnamefont {A.}~\bibnamefont {Maiezza}},
  \ and\ \bibinfo {author} {\bibfnamefont {F.}~\bibnamefont {Nesti}},\ }\href
  {\doibase 10.1103/PhysRevD.89.095028} {\bibfield  {journal} {\bibinfo
  {journal} {Phys. Rev. D}\ }\textbf {\bibinfo {volume} {89}},\ \bibinfo
  {pages} {095028} (\bibinfo {year} {2014})},\ \Eprint
  {http://arxiv.org/abs/1403.7112} {arXiv:1403.7112 [hep-ph]} \BibitemShut
  {NoStop}%
\bibitem [{\citenamefont {Bertolini}\ \emph {et~al.}(2020)\citenamefont
  {Bertolini}, \citenamefont {Maiezza},\ and\ \citenamefont
  {Nesti}}]{Bertolini:2019out}%
  \BibitemOpen
  \bibfield  {author} {\bibinfo {author} {\bibfnamefont {S.}~\bibnamefont
  {Bertolini}}, \bibinfo {author} {\bibfnamefont {A.}~\bibnamefont {Maiezza}},
  \ and\ \bibinfo {author} {\bibfnamefont {F.}~\bibnamefont {Nesti}},\ }\href
  {\doibase 10.1103/PhysRevD.101.035036} {\bibfield  {journal} {\bibinfo
  {journal} {Phys. Rev. D}\ }\textbf {\bibinfo {volume} {101}},\ \bibinfo
  {pages} {035036} (\bibinfo {year} {2020})},\ \Eprint
  {http://arxiv.org/abs/1911.09472} {arXiv:1911.09472 [hep-ph]} \BibitemShut
  {NoStop}%
\bibitem [{\citenamefont {Adams}\ \emph {et~al.}(2022)\citenamefont {Adams}
  \emph {et~al.}}]{Adams:2022jwx}%
  \BibitemOpen
  \bibfield  {author} {\bibinfo {author} {\bibfnamefont {C.}~\bibnamefont
  {Adams}} \emph {et~al.},\ }\href@noop {} {\  (\bibinfo {year} {2022})},\
  \Eprint {http://arxiv.org/abs/2212.11099} {arXiv:2212.11099 [nucl-ex]}
  \BibitemShut {NoStop}%
\bibitem [{\citenamefont {Bhupal~Dev}\ \emph {et~al.}(2013)\citenamefont
  {Bhupal~Dev}, \citenamefont {Goswami}, \citenamefont {Mitra},\ and\
  \citenamefont {Rodejohann}}]{BhupalDev:2013ntw}%
  \BibitemOpen
  \bibfield  {author} {\bibinfo {author} {\bibfnamefont {P.~S.}\ \bibnamefont
  {Bhupal~Dev}}, \bibinfo {author} {\bibfnamefont {S.}~\bibnamefont {Goswami}},
  \bibinfo {author} {\bibfnamefont {M.}~\bibnamefont {Mitra}}, \ and\ \bibinfo
  {author} {\bibfnamefont {W.}~\bibnamefont {Rodejohann}},\ }\href {\doibase
  10.1103/PhysRevD.88.091301} {\bibfield  {journal} {\bibinfo  {journal} {Phys.
  Rev. D}\ }\textbf {\bibinfo {volume} {88}},\ \bibinfo {pages} {091301}
  (\bibinfo {year} {2013})},\ \Eprint {http://arxiv.org/abs/1305.0056}
  {arXiv:1305.0056 [hep-ph]} \BibitemShut {NoStop}%
\bibitem [{\citenamefont {Cirigliano}\ \emph
  {et~al.}(2018{\natexlab{a}})\citenamefont {Cirigliano}, \citenamefont
  {Dekens}, \citenamefont {de~Vries}, \citenamefont {Graesser},\ and\
  \citenamefont {Mereghetti}}]{Cirigliano:2018yza}%
  \BibitemOpen
  \bibfield  {author} {\bibinfo {author} {\bibfnamefont {V.}~\bibnamefont
  {Cirigliano}}, \bibinfo {author} {\bibfnamefont {W.}~\bibnamefont {Dekens}},
  \bibinfo {author} {\bibfnamefont {J.}~\bibnamefont {de~Vries}}, \bibinfo
  {author} {\bibfnamefont {M.~L.}\ \bibnamefont {Graesser}}, \ and\ \bibinfo
  {author} {\bibfnamefont {E.}~\bibnamefont {Mereghetti}},\ }\href {\doibase
  10.1007/JHEP12(2018)097} {\bibfield  {journal} {\bibinfo  {journal} {JHEP}\
  }\textbf {\bibinfo {volume} {12}},\ \bibinfo {pages} {097} (\bibinfo {year}
  {2018}{\natexlab{a}})},\ \Eprint {http://arxiv.org/abs/1806.02780}
  {arXiv:1806.02780 [hep-ph]} \BibitemShut {NoStop}%
\bibitem [{\citenamefont {Horoi}\ and\ \citenamefont
  {Neacsu}(2017)}]{Horoi:2017gmj}%
  \BibitemOpen
  \bibfield  {author} {\bibinfo {author} {\bibfnamefont {M.}~\bibnamefont
  {Horoi}}\ and\ \bibinfo {author} {\bibfnamefont {A.}~\bibnamefont {Neacsu}},\
  }\href@noop {} {\  (\bibinfo {year} {2017})},\ \Eprint
  {http://arxiv.org/abs/1706.05391} {arXiv:1706.05391 [hep-ph]} \BibitemShut
  {NoStop}%
\bibitem [{\citenamefont {Dekens}\ \emph {et~al.}(2020)\citenamefont {Dekens},
  \citenamefont {de~Vries}, \citenamefont {Fuyuto}, \citenamefont
  {Mereghetti},\ and\ \citenamefont {Zhou}}]{Dekens:2020ttz}%
  \BibitemOpen
  \bibfield  {author} {\bibinfo {author} {\bibfnamefont {W.}~\bibnamefont
  {Dekens}}, \bibinfo {author} {\bibfnamefont {J.}~\bibnamefont {de~Vries}},
  \bibinfo {author} {\bibfnamefont {K.}~\bibnamefont {Fuyuto}}, \bibinfo
  {author} {\bibfnamefont {E.}~\bibnamefont {Mereghetti}}, \ and\ \bibinfo
  {author} {\bibfnamefont {G.}~\bibnamefont {Zhou}},\ }\href {\doibase
  10.1007/JHEP06(2020)097} {\bibfield  {journal} {\bibinfo  {journal} {JHEP}\
  }\textbf {\bibinfo {volume} {06}},\ \bibinfo {pages} {097} (\bibinfo {year}
  {2020})},\ \Eprint {http://arxiv.org/abs/2002.07182} {arXiv:2002.07182
  [hep-ph]} \BibitemShut {NoStop}%
\bibitem [{\citenamefont {Ding}\ \emph {et~al.}(2024)\citenamefont {Ding},
  \citenamefont {Li},\ and\ \citenamefont {Yao}}]{Ding:2024obt}%
  \BibitemOpen
  \bibfield  {author} {\bibinfo {author} {\bibfnamefont {C.-r.}\ \bibnamefont
  {Ding}}, \bibinfo {author} {\bibfnamefont {G.}~\bibnamefont {Li}}, \ and\
  \bibinfo {author} {\bibfnamefont {J.-m.}\ \bibnamefont {Yao}},\ }\href
  {\doibase 10.1016/j.physletb.2024.138896} {\bibfield  {journal} {\bibinfo
  {journal} {Phys. Lett. B}\ }\textbf {\bibinfo {volume} {856}},\ \bibinfo
  {pages} {138896} (\bibinfo {year} {2024})},\ \Eprint
  {http://arxiv.org/abs/2403.17722} {arXiv:2403.17722 [nucl-th]} \BibitemShut
  {NoStop}%
\bibitem [{\citenamefont {Doi}\ \emph {et~al.}(1985)\citenamefont {Doi},
  \citenamefont {Kotani},\ and\ \citenamefont {Takasugi}}]{Doi:1985dx}%
  \BibitemOpen
  \bibfield  {author} {\bibinfo {author} {\bibfnamefont {M.}~\bibnamefont
  {Doi}}, \bibinfo {author} {\bibfnamefont {T.}~\bibnamefont {Kotani}}, \ and\
  \bibinfo {author} {\bibfnamefont {E.}~\bibnamefont {Takasugi}},\ }\href
  {\doibase 10.1143/PTPS.83.1} {\bibfield  {journal} {\bibinfo  {journal}
  {Prog. Theor. Phys. Suppl.}\ }\textbf {\bibinfo {volume} {83}},\ \bibinfo
  {pages} {1} (\bibinfo {year} {1985})}\BibitemShut {NoStop}%
\bibitem [{\citenamefont {Hirsch}\ \emph {et~al.}(1996)\citenamefont {Hirsch},
  \citenamefont {Klapdor-Kleingrothaus},\ and\ \citenamefont
  {Panella}}]{Hirsch:1996qw}%
  \BibitemOpen
  \bibfield  {author} {\bibinfo {author} {\bibfnamefont {M.}~\bibnamefont
  {Hirsch}}, \bibinfo {author} {\bibfnamefont {H.~V.}\ \bibnamefont
  {Klapdor-Kleingrothaus}}, \ and\ \bibinfo {author} {\bibfnamefont
  {O.}~\bibnamefont {Panella}},\ }\href {\doibase 10.1016/0370-2693(96)00185-2}
  {\bibfield  {journal} {\bibinfo  {journal} {Phys. Lett. B}\ }\textbf
  {\bibinfo {volume} {374}},\ \bibinfo {pages} {7} (\bibinfo {year} {1996})},\
  \Eprint {http://arxiv.org/abs/hep-ph/9602306} {arXiv:hep-ph/9602306}
  \BibitemShut {NoStop}%
\bibitem [{\citenamefont {Vergados}\ \emph {et~al.}(2012)\citenamefont
  {Vergados}, \citenamefont {Ejiri},\ and\ \citenamefont
  {Simkovic}}]{Vergados:2012xy}%
  \BibitemOpen
  \bibfield  {author} {\bibinfo {author} {\bibfnamefont {J.~D.}\ \bibnamefont
  {Vergados}}, \bibinfo {author} {\bibfnamefont {H.}~\bibnamefont {Ejiri}}, \
  and\ \bibinfo {author} {\bibfnamefont {F.}~\bibnamefont {Simkovic}},\ }\href
  {\doibase 10.1088/0034-4885/75/10/106301} {\bibfield  {journal} {\bibinfo
  {journal} {Rept. Prog. Phys.}\ }\textbf {\bibinfo {volume} {75}},\ \bibinfo
  {pages} {106301} (\bibinfo {year} {2012})},\ \Eprint
  {http://arxiv.org/abs/1205.0649} {arXiv:1205.0649 [hep-ph]} \BibitemShut
  {NoStop}%
\bibitem [{\citenamefont {Dekens}\ \emph {et~al.}(2024)\citenamefont {Dekens},
  \citenamefont {de~Vries}, \citenamefont {Castillo}, \citenamefont
  {Men{\'e}ndez}, \citenamefont {Mereghetti}, \citenamefont {Plakkot},
  \citenamefont {Soriano},\ and\ \citenamefont {Zhou}}]{Dekens:2024hlz}%
  \BibitemOpen
  \bibfield  {author} {\bibinfo {author} {\bibfnamefont {W.}~\bibnamefont
  {Dekens}}, \bibinfo {author} {\bibfnamefont {J.}~\bibnamefont {de~Vries}},
  \bibinfo {author} {\bibfnamefont {D.}~\bibnamefont {Castillo}}, \bibinfo
  {author} {\bibfnamefont {J.}~\bibnamefont {Men{\'e}ndez}}, \bibinfo {author}
  {\bibfnamefont {E.}~\bibnamefont {Mereghetti}}, \bibinfo {author}
  {\bibfnamefont {V.}~\bibnamefont {Plakkot}}, \bibinfo {author} {\bibfnamefont
  {P.}~\bibnamefont {Soriano}}, \ and\ \bibinfo {author} {\bibfnamefont
  {G.}~\bibnamefont {Zhou}},\ }\href {\doibase 10.1007/JHEP09(2024)201}
  {\bibfield  {journal} {\bibinfo  {journal} {JHEP}\ }\textbf {\bibinfo
  {volume} {09}},\ \bibinfo {pages} {201} (\bibinfo {year} {2024})},\ \Eprint
  {http://arxiv.org/abs/2402.07993} {arXiv:2402.07993 [hep-ph]} \BibitemShut
  {NoStop}%
\bibitem [{\citenamefont {Jokiniemi}\ \emph {et~al.}(2021)\citenamefont
  {Jokiniemi}, \citenamefont {Soriano},\ and\ \citenamefont
  {Men{\'e}ndez}}]{Jokiniemi:2021qqv}%
  \BibitemOpen
  \bibfield  {author} {\bibinfo {author} {\bibfnamefont {L.}~\bibnamefont
  {Jokiniemi}}, \bibinfo {author} {\bibfnamefont {P.}~\bibnamefont {Soriano}},
  \ and\ \bibinfo {author} {\bibfnamefont {J.}~\bibnamefont {Men{\'e}ndez}},\
  }\href {\doibase 10.1016/j.physletb.2021.136720} {\bibfield  {journal}
  {\bibinfo  {journal} {Phys. Lett. B}\ }\textbf {\bibinfo {volume} {823}},\
  \bibinfo {pages} {136720} (\bibinfo {year} {2021})},\ \Eprint
  {http://arxiv.org/abs/2107.13354} {arXiv:2107.13354 [nucl-th]} \BibitemShut
  {NoStop}%
\bibitem [{\citenamefont {Cirigliano}\ \emph
  {et~al.}(2018{\natexlab{b}})\citenamefont {Cirigliano}, \citenamefont
  {Dekens}, \citenamefont {De~Vries}, \citenamefont {Graesser}, \citenamefont
  {Mereghetti}, \citenamefont {Pastore},\ and\ \citenamefont
  {Van~Kolck}}]{Cirigliano:2018hja}%
  \BibitemOpen
  \bibfield  {author} {\bibinfo {author} {\bibfnamefont {V.}~\bibnamefont
  {Cirigliano}}, \bibinfo {author} {\bibfnamefont {W.}~\bibnamefont {Dekens}},
  \bibinfo {author} {\bibfnamefont {J.}~\bibnamefont {De~Vries}}, \bibinfo
  {author} {\bibfnamefont {M.~L.}\ \bibnamefont {Graesser}}, \bibinfo {author}
  {\bibfnamefont {E.}~\bibnamefont {Mereghetti}}, \bibinfo {author}
  {\bibfnamefont {S.}~\bibnamefont {Pastore}}, \ and\ \bibinfo {author}
  {\bibfnamefont {U.}~\bibnamefont {Van~Kolck}},\ }\href {\doibase
  10.1103/PhysRevLett.120.202001} {\bibfield  {journal} {\bibinfo  {journal}
  {Phys. Rev. Lett.}\ }\textbf {\bibinfo {volume} {120}},\ \bibinfo {pages}
  {202001} (\bibinfo {year} {2018}{\natexlab{b}})},\ \Eprint
  {http://arxiv.org/abs/1802.10097} {arXiv:1802.10097 [hep-ph]} \BibitemShut
  {NoStop}%
\bibitem [{\citenamefont {Cirigliano}\ \emph {et~al.}(2019)\citenamefont
  {Cirigliano}, \citenamefont {Dekens}, \citenamefont {De~Vries}, \citenamefont
  {Graesser}, \citenamefont {Mereghetti}, \citenamefont {Pastore},
  \citenamefont {Piarulli}, \citenamefont {Van~Kolck},\ and\ \citenamefont
  {Wiringa}}]{Cirigliano:2019vdj}%
  \BibitemOpen
  \bibfield  {author} {\bibinfo {author} {\bibfnamefont {V.}~\bibnamefont
  {Cirigliano}}, \bibinfo {author} {\bibfnamefont {W.}~\bibnamefont {Dekens}},
  \bibinfo {author} {\bibfnamefont {J.}~\bibnamefont {De~Vries}}, \bibinfo
  {author} {\bibfnamefont {M.~L.}\ \bibnamefont {Graesser}}, \bibinfo {author}
  {\bibfnamefont {E.}~\bibnamefont {Mereghetti}}, \bibinfo {author}
  {\bibfnamefont {S.}~\bibnamefont {Pastore}}, \bibinfo {author} {\bibfnamefont
  {M.}~\bibnamefont {Piarulli}}, \bibinfo {author} {\bibfnamefont
  {U.}~\bibnamefont {Van~Kolck}}, \ and\ \bibinfo {author} {\bibfnamefont
  {R.~B.}\ \bibnamefont {Wiringa}},\ }\href {\doibase
  10.1103/PhysRevC.100.055504} {\bibfield  {journal} {\bibinfo  {journal}
  {Phys. Rev. C}\ }\textbf {\bibinfo {volume} {100}},\ \bibinfo {pages}
  {055504} (\bibinfo {year} {2019})},\ \Eprint
  {http://arxiv.org/abs/1907.11254} {arXiv:1907.11254 [nucl-th]} \BibitemShut
  {NoStop}%
\bibitem [{\citenamefont {Cirigliano}\ \emph
  {et~al.}(2021{\natexlab{a}})\citenamefont {Cirigliano}, \citenamefont
  {Dekens}, \citenamefont {de~Vries}, \citenamefont {Hoferichter},\ and\
  \citenamefont {Mereghetti}}]{Cirigliano:2020dmx}%
  \BibitemOpen
  \bibfield  {author} {\bibinfo {author} {\bibfnamefont {V.}~\bibnamefont
  {Cirigliano}}, \bibinfo {author} {\bibfnamefont {W.}~\bibnamefont {Dekens}},
  \bibinfo {author} {\bibfnamefont {J.}~\bibnamefont {de~Vries}}, \bibinfo
  {author} {\bibfnamefont {M.}~\bibnamefont {Hoferichter}}, \ and\ \bibinfo
  {author} {\bibfnamefont {E.}~\bibnamefont {Mereghetti}},\ }\href {\doibase
  10.1103/PhysRevLett.126.172002} {\bibfield  {journal} {\bibinfo  {journal}
  {Phys. Rev. Lett.}\ }\textbf {\bibinfo {volume} {126}},\ \bibinfo {pages}
  {172002} (\bibinfo {year} {2021}{\natexlab{a}})},\ \Eprint
  {http://arxiv.org/abs/2012.11602} {arXiv:2012.11602 [nucl-th]} \BibitemShut
  {NoStop}%
\bibitem [{\citenamefont {Cirigliano}\ \emph
  {et~al.}(2021{\natexlab{b}})\citenamefont {Cirigliano}, \citenamefont
  {Dekens}, \citenamefont {de~Vries}, \citenamefont {Hoferichter},\ and\
  \citenamefont {Mereghetti}}]{Cirigliano:2021qko}%
  \BibitemOpen
  \bibfield  {author} {\bibinfo {author} {\bibfnamefont {V.}~\bibnamefont
  {Cirigliano}}, \bibinfo {author} {\bibfnamefont {W.}~\bibnamefont {Dekens}},
  \bibinfo {author} {\bibfnamefont {J.}~\bibnamefont {de~Vries}}, \bibinfo
  {author} {\bibfnamefont {M.}~\bibnamefont {Hoferichter}}, \ and\ \bibinfo
  {author} {\bibfnamefont {E.}~\bibnamefont {Mereghetti}},\ }\href {\doibase
  10.1007/JHEP05(2021)289} {\bibfield  {journal} {\bibinfo  {journal} {JHEP}\
  }\textbf {\bibinfo {volume} {05}},\ \bibinfo {pages} {289} (\bibinfo {year}
  {2021}{\natexlab{b}})},\ \Eprint {http://arxiv.org/abs/2102.03371}
  {arXiv:2102.03371 [nucl-th]} \BibitemShut {NoStop}%
\bibitem [{\citenamefont {Weinberg}(1990)}]{Weinberg:1990rz}%
  \BibitemOpen
  \bibfield  {author} {\bibinfo {author} {\bibfnamefont {S.}~\bibnamefont
  {Weinberg}},\ }\href {\doibase 10.1016/0370-2693(90)90938-3} {\bibfield
  {journal} {\bibinfo  {journal} {Phys. Lett. B}\ }\textbf {\bibinfo {volume}
  {251}},\ \bibinfo {pages} {288} (\bibinfo {year} {1990})}\BibitemShut
  {NoStop}%
\bibitem [{\citenamefont {Weinberg}(1991)}]{Weinberg:1991um}%
  \BibitemOpen
  \bibfield  {author} {\bibinfo {author} {\bibfnamefont {S.}~\bibnamefont
  {Weinberg}},\ }\href {\doibase 10.1016/0550-3213(91)90231-L} {\bibfield
  {journal} {\bibinfo  {journal} {Nucl. Phys. B}\ }\textbf {\bibinfo {volume}
  {363}},\ \bibinfo {pages} {3} (\bibinfo {year} {1991})}\BibitemShut {NoStop}%
\bibitem [{\citenamefont {Wirth}\ \emph {et~al.}(2021)\citenamefont {Wirth},
  \citenamefont {Yao},\ and\ \citenamefont {Hergert}}]{Wirth:2021pij}%
  \BibitemOpen
  \bibfield  {author} {\bibinfo {author} {\bibfnamefont {R.}~\bibnamefont
  {Wirth}}, \bibinfo {author} {\bibfnamefont {J.~M.}\ \bibnamefont {Yao}}, \
  and\ \bibinfo {author} {\bibfnamefont {H.}~\bibnamefont {Hergert}},\ }\href
  {\doibase 10.1103/PhysRevLett.127.242502} {\bibfield  {journal} {\bibinfo
  {journal} {Phys. Rev. Lett.}\ }\textbf {\bibinfo {volume} {127}},\ \bibinfo
  {pages} {242502} (\bibinfo {year} {2021})},\ \Eprint
  {http://arxiv.org/abs/2105.05415} {arXiv:2105.05415 [nucl-th]} \BibitemShut
  {NoStop}%
\bibitem [{\citenamefont {Men{\'e}ndez}(2018)}]{Menendez:2017fdf}%
  \BibitemOpen
  \bibfield  {author} {\bibinfo {author} {\bibfnamefont {J.}~\bibnamefont
  {Men{\'e}ndez}},\ }\href {\doibase 10.1088/1361-6471/aa9bd4} {\bibfield
  {journal} {\bibinfo  {journal} {J. Phys. G}\ }\textbf {\bibinfo {volume}
  {45}},\ \bibinfo {pages} {014003} (\bibinfo {year} {2018})},\ \Eprint
  {http://arxiv.org/abs/1804.02105} {arXiv:1804.02105 [nucl-th]} \BibitemShut
  {NoStop}%
\bibitem [{\citenamefont {Hyv{\"a}rinen}\ and\ \citenamefont
  {Suhonen}(2015)}]{Hyvarinen:2015bda}%
  \BibitemOpen
  \bibfield  {author} {\bibinfo {author} {\bibfnamefont {J.}~\bibnamefont
  {Hyv{\"a}rinen}}\ and\ \bibinfo {author} {\bibfnamefont {J.}~\bibnamefont
  {Suhonen}},\ }\href {\doibase 10.1103/PhysRevC.91.024613} {\bibfield
  {journal} {\bibinfo  {journal} {Phys. Rev. C}\ }\textbf {\bibinfo {volume}
  {91}},\ \bibinfo {pages} {024613} (\bibinfo {year} {2015})}\BibitemShut
  {NoStop}%
\bibitem [{\citenamefont {Gr{\'a}f}\ \emph {et~al.}(2022)\citenamefont
  {Gr{\'a}f}, \citenamefont {Lindner},\ and\ \citenamefont
  {Scholer}}]{Graf:2022lhj}%
  \BibitemOpen
  \bibfield  {author} {\bibinfo {author} {\bibfnamefont {L.}~\bibnamefont
  {Gr{\'a}f}}, \bibinfo {author} {\bibfnamefont {M.}~\bibnamefont {Lindner}}, \
  and\ \bibinfo {author} {\bibfnamefont {O.}~\bibnamefont {Scholer}},\ }\href
  {\doibase 10.1103/PhysRevD.106.035022} {\bibfield  {journal} {\bibinfo
  {journal} {Phys. Rev. D}\ }\textbf {\bibinfo {volume} {106}},\ \bibinfo
  {pages} {035022} (\bibinfo {year} {2022})},\ \Eprint
  {http://arxiv.org/abs/2204.10845} {arXiv:2204.10845 [hep-ph]} \BibitemShut
  {NoStop}%
\bibitem [{\citenamefont {Bollig}\ \emph {et~al.}(2020)\citenamefont {Bollig},
  \citenamefont {DeRocco}, \citenamefont {Graham},\ and\ \citenamefont
  {Janka}}]{Bollig:2020xdr}%
  \BibitemOpen
  \bibfield  {author} {\bibinfo {author} {\bibfnamefont {R.}~\bibnamefont
  {Bollig}}, \bibinfo {author} {\bibfnamefont {W.}~\bibnamefont {DeRocco}},
  \bibinfo {author} {\bibfnamefont {P.~W.}\ \bibnamefont {Graham}}, \ and\
  \bibinfo {author} {\bibfnamefont {H.-T.}\ \bibnamefont {Janka}},\ }\href
  {\doibase 10.1103/PhysRevLett.125.051104} {\bibfield  {journal} {\bibinfo
  {journal} {Phys. Rev. Lett.}\ }\textbf {\bibinfo {volume} {125}},\ \bibinfo
  {pages} {051104} (\bibinfo {year} {2020})},\ \bibinfo {note} {[Erratum:
  Phys.Rev.Lett. 126, 189901 (2021)]. Data can be found from
  \url{https://wwwmpa.mpa-garching.mpg.de/ccsnarchive/data/Bollig2016_radial_profiles/}.},\
  \Eprint {http://arxiv.org/abs/2005.07141} {arXiv:2005.07141 [hep-ph]}
  \BibitemShut {NoStop}%
\bibitem [{\citenamefont {Carenza}\ \emph {et~al.}(2024)\citenamefont
  {Carenza}, \citenamefont {Lucente}, \citenamefont {Mastrototaro},
  \citenamefont {Mirizzi},\ and\ \citenamefont {Serpico}}]{Carenza:2023old}%
  \BibitemOpen
  \bibfield  {author} {\bibinfo {author} {\bibfnamefont {P.}~\bibnamefont
  {Carenza}}, \bibinfo {author} {\bibfnamefont {G.}~\bibnamefont {Lucente}},
  \bibinfo {author} {\bibfnamefont {L.}~\bibnamefont {Mastrototaro}}, \bibinfo
  {author} {\bibfnamefont {A.}~\bibnamefont {Mirizzi}}, \ and\ \bibinfo
  {author} {\bibfnamefont {P.~D.}\ \bibnamefont {Serpico}},\ }\href {\doibase
  10.1103/PhysRevD.109.063010} {\bibfield  {journal} {\bibinfo  {journal}
  {Phys. Rev. D}\ }\textbf {\bibinfo {volume} {109}},\ \bibinfo {pages}
  {063010} (\bibinfo {year} {2024})},\ \Eprint
  {http://arxiv.org/abs/2311.00033} {arXiv:2311.00033 [hep-ph]} \BibitemShut
  {NoStop}%
\bibitem [{\citenamefont {Chauhan}\ \emph {et~al.}(2025)\citenamefont
  {Chauhan}, \citenamefont {Gustafson},\ and\ \citenamefont
  {Shoemaker}}]{Chauhan:2025mnn}%
  \BibitemOpen
  \bibfield  {author} {\bibinfo {author} {\bibfnamefont {G.}~\bibnamefont
  {Chauhan}}, \bibinfo {author} {\bibfnamefont {R.~A.}\ \bibnamefont
  {Gustafson}}, \ and\ \bibinfo {author} {\bibfnamefont {I.~M.}\ \bibnamefont
  {Shoemaker}},\ }\href@noop {} {\  (\bibinfo {year} {2025})},\ \Eprint
  {http://arxiv.org/abs/2503.13607} {arXiv:2503.13607 [hep-ph]} \BibitemShut
  {NoStop}%
\bibitem [{\citenamefont {Fiorillo}\ \emph {et~al.}(2025)\citenamefont
  {Fiorillo}, \citenamefont {Pitik},\ and\ \citenamefont
  {Vitagliano}}]{Fiorillo:2025yzf}%
  \BibitemOpen
  \bibfield  {author} {\bibinfo {author} {\bibfnamefont {D.~F.~G.}\
  \bibnamefont {Fiorillo}}, \bibinfo {author} {\bibfnamefont {T.}~\bibnamefont
  {Pitik}}, \ and\ \bibinfo {author} {\bibfnamefont {E.}~\bibnamefont
  {Vitagliano}},\ }\href {\doibase 10.1103/cz94-dqxt} {\bibfield  {journal}
  {\bibinfo  {journal} {Phys. Rev. Lett.}\ }\textbf {\bibinfo {volume} {135}},\
  \bibinfo {pages} {071005} (\bibinfo {year} {2025})},\ \Eprint
  {http://arxiv.org/abs/2503.13653} {arXiv:2503.13653 [hep-ph]} \BibitemShut
  {NoStop}%
\bibitem [{\citenamefont {Caputo}\ \emph {et~al.}(2022)\citenamefont {Caputo},
  \citenamefont {Janka}, \citenamefont {Raffelt},\ and\ \citenamefont
  {Vitagliano}}]{Caputo:2022mah}%
  \BibitemOpen
  \bibfield  {author} {\bibinfo {author} {\bibfnamefont {A.}~\bibnamefont
  {Caputo}}, \bibinfo {author} {\bibfnamefont {H.-T.}\ \bibnamefont {Janka}},
  \bibinfo {author} {\bibfnamefont {G.}~\bibnamefont {Raffelt}}, \ and\
  \bibinfo {author} {\bibfnamefont {E.}~\bibnamefont {Vitagliano}},\ }\href
  {\doibase 10.1103/PhysRevLett.128.221103} {\bibfield  {journal} {\bibinfo
  {journal} {Phys. Rev. Lett.}\ }\textbf {\bibinfo {volume} {128}},\ \bibinfo
  {pages} {221103} (\bibinfo {year} {2022})},\ \Eprint
  {http://arxiv.org/abs/2201.09890} {arXiv:2201.09890 [astro-ph.HE]}
  \BibitemShut {NoStop}%
\bibitem [{\citenamefont {Sung}\ \emph {et~al.}(2019)\citenamefont {Sung},
  \citenamefont {Tu},\ and\ \citenamefont {Wu}}]{Sung:2019xie}%
  \BibitemOpen
  \bibfield  {author} {\bibinfo {author} {\bibfnamefont {A.}~\bibnamefont
  {Sung}}, \bibinfo {author} {\bibfnamefont {H.}~\bibnamefont {Tu}}, \ and\
  \bibinfo {author} {\bibfnamefont {M.-R.}\ \bibnamefont {Wu}},\ }\href
  {\doibase 10.1103/PhysRevD.99.121305} {\bibfield  {journal} {\bibinfo
  {journal} {Phys. Rev. D}\ }\textbf {\bibinfo {volume} {99}},\ \bibinfo
  {pages} {121305} (\bibinfo {year} {2019})},\ \Eprint
  {http://arxiv.org/abs/1903.07923} {arXiv:1903.07923 [hep-ph]} \BibitemShut
  {NoStop}%
\bibitem [{\citenamefont {Falk}\ and\ \citenamefont
  {Schramm}(1978)}]{Falk:1978kf}%
  \BibitemOpen
  \bibfield  {author} {\bibinfo {author} {\bibfnamefont {S.~W.}\ \bibnamefont
  {Falk}}\ and\ \bibinfo {author} {\bibfnamefont {D.~N.}\ \bibnamefont
  {Schramm}},\ }\href {\doibase 10.1016/0370-2693(78)90417-3} {\bibfield
  {journal} {\bibinfo  {journal} {Phys. Lett. B}\ }\textbf {\bibinfo {volume}
  {79}},\ \bibinfo {pages} {511} (\bibinfo {year} {1978})}\BibitemShut
  {NoStop}%
\bibitem [{\citenamefont {Rembiasz}\ \emph {et~al.}(2018)\citenamefont
  {Rembiasz}, \citenamefont {Obergaulinger}, \citenamefont {Masip},
  \citenamefont {P{\'e}rez-Garc{\'\i}a}, \citenamefont {Aloy},\ and\
  \citenamefont {Albertus}}]{Rembiasz:2018lok}%
  \BibitemOpen
  \bibfield  {author} {\bibinfo {author} {\bibfnamefont {T.}~\bibnamefont
  {Rembiasz}}, \bibinfo {author} {\bibfnamefont {M.}~\bibnamefont
  {Obergaulinger}}, \bibinfo {author} {\bibfnamefont {M.}~\bibnamefont
  {Masip}}, \bibinfo {author} {\bibfnamefont {M.~{\'A}.}\ \bibnamefont
  {P{\'e}rez-Garc{\'\i}a}}, \bibinfo {author} {\bibfnamefont {M.-{\'A}.}\
  \bibnamefont {Aloy}}, \ and\ \bibinfo {author} {\bibfnamefont
  {C.}~\bibnamefont {Albertus}},\ }\href {\doibase 10.1103/PhysRevD.98.103010}
  {\bibfield  {journal} {\bibinfo  {journal} {Phys. Rev. D}\ }\textbf {\bibinfo
  {volume} {98}},\ \bibinfo {pages} {103010} (\bibinfo {year} {2018})},\
  \Eprint {http://arxiv.org/abs/1806.03300} {arXiv:1806.03300 [astro-ph.HE]}
  \BibitemShut {NoStop}%
\bibitem [{\citenamefont {Hirata}\ \emph {et~al.}(1987)\citenamefont {Hirata}
  \emph {et~al.}}]{Kamiokande-II:1987idp}%
  \BibitemOpen
  \bibfield  {author} {\bibinfo {author} {\bibfnamefont {K.}~\bibnamefont
  {Hirata}} \emph {et~al.} (\bibinfo {collaboration} {Kamiokande-II}),\ }\href
  {\doibase 10.1103/PhysRevLett.58.1490} {\bibfield  {journal} {\bibinfo
  {journal} {Phys. Rev. Lett.}\ }\textbf {\bibinfo {volume} {58}},\ \bibinfo
  {pages} {1490} (\bibinfo {year} {1987})}\BibitemShut {NoStop}%
\bibitem [{\citenamefont {Bionta}\ \emph {et~al.}(1987)\citenamefont {Bionta}
  \emph {et~al.}}]{Bionta:1987qt}%
  \BibitemOpen
  \bibfield  {author} {\bibinfo {author} {\bibfnamefont {R.~M.}\ \bibnamefont
  {Bionta}} \emph {et~al.},\ }\href {\doibase 10.1103/PhysRevLett.58.1494}
  {\bibfield  {journal} {\bibinfo  {journal} {Phys. Rev. Lett.}\ }\textbf
  {\bibinfo {volume} {58}},\ \bibinfo {pages} {1494} (\bibinfo {year}
  {1987})}\BibitemShut {NoStop}%
\bibitem [{\citenamefont {Alekseev}\ \emph {et~al.}(1988)\citenamefont
  {Alekseev}, \citenamefont {Alekseeva}, \citenamefont {Krivosheina},\ and\
  \citenamefont {Volchenko}}]{Alekseev:1988gp}%
  \BibitemOpen
  \bibfield  {author} {\bibinfo {author} {\bibfnamefont {E.~N.}\ \bibnamefont
  {Alekseev}}, \bibinfo {author} {\bibfnamefont {L.~N.}\ \bibnamefont
  {Alekseeva}}, \bibinfo {author} {\bibfnamefont {I.~V.}\ \bibnamefont
  {Krivosheina}}, \ and\ \bibinfo {author} {\bibfnamefont {V.~I.}\ \bibnamefont
  {Volchenko}},\ }\href {\doibase 10.1016/0370-2693(88)91651-6} {\bibfield
  {journal} {\bibinfo  {journal} {Phys. Lett. B}\ }\textbf {\bibinfo {volume}
  {205}},\ \bibinfo {pages} {209} (\bibinfo {year} {1988})}\BibitemShut
  {NoStop}%
\bibitem [{\citenamefont {Raffelt}(1996)}]{raffelt1996stars}%
  \BibitemOpen
  \bibfield  {author} {\bibinfo {author} {\bibfnamefont {G.~G.}\ \bibnamefont
  {Raffelt}},\ }\href@noop {} {\emph {\bibinfo {title} {{Stars as laboratories
  for fundamental physics: The astrophysics of neutrinos, axions, and other
  weakly interacting particles}}}}\ (\bibinfo  {publisher} {University of
  Chicago press},\ \bibinfo {year} {1996})\BibitemShut {NoStop}%
\bibitem [{\citenamefont {Dreiner}\ \emph {et~al.}(2003)\citenamefont
  {Dreiner}, \citenamefont {Hanhart}, \citenamefont {Langenfeld},\ and\
  \citenamefont {Phillips}}]{Dreiner:2003wh}%
  \BibitemOpen
  \bibfield  {author} {\bibinfo {author} {\bibfnamefont {H.~K.}\ \bibnamefont
  {Dreiner}}, \bibinfo {author} {\bibfnamefont {C.}~\bibnamefont {Hanhart}},
  \bibinfo {author} {\bibfnamefont {U.}~\bibnamefont {Langenfeld}}, \ and\
  \bibinfo {author} {\bibfnamefont {D.~R.}\ \bibnamefont {Phillips}},\ }\href
  {\doibase 10.1103/PhysRevD.68.055004} {\bibfield  {journal} {\bibinfo
  {journal} {Phys. Rev. D}\ }\textbf {\bibinfo {volume} {68}},\ \bibinfo
  {pages} {055004} (\bibinfo {year} {2003})},\ \Eprint
  {http://arxiv.org/abs/hep-ph/0304289} {arXiv:hep-ph/0304289} \BibitemShut
  {NoStop}%
\bibitem [{\citenamefont {Nemevsek}\ \emph {et~al.}(2011)\citenamefont
  {Nemevsek}, \citenamefont {Nesti}, \citenamefont {Senjanovic},\ and\
  \citenamefont {Tello}}]{Nemevsek:2011aa}%
  \BibitemOpen
  \bibfield  {author} {\bibinfo {author} {\bibfnamefont {M.}~\bibnamefont
  {Nemevsek}}, \bibinfo {author} {\bibfnamefont {F.}~\bibnamefont {Nesti}},
  \bibinfo {author} {\bibfnamefont {G.}~\bibnamefont {Senjanovic}}, \ and\
  \bibinfo {author} {\bibfnamefont {V.}~\bibnamefont {Tello}},\ }\href@noop {}
  {\  (\bibinfo {year} {2011})},\ \Eprint {http://arxiv.org/abs/1112.3061}
  {arXiv:1112.3061 [hep-ph]} \BibitemShut {NoStop}%
\bibitem [{\citenamefont {Nemev{\v{s}}ek}\ and\ \citenamefont
  {Nesti}(2023)}]{Nemevsek:2023hwx}%
  \BibitemOpen
  \bibfield  {author} {\bibinfo {author} {\bibfnamefont {M.}~\bibnamefont
  {Nemev{\v{s}}ek}}\ and\ \bibinfo {author} {\bibfnamefont {F.}~\bibnamefont
  {Nesti}},\ }\href {\doibase 10.1103/PhysRevD.108.015030} {\bibfield
  {journal} {\bibinfo  {journal} {Phys. Rev. D}\ }\textbf {\bibinfo {volume}
  {108}},\ \bibinfo {pages} {015030} (\bibinfo {year} {2023})},\ \Eprint
  {http://arxiv.org/abs/2306.12104} {arXiv:2306.12104 [hep-ph]} \BibitemShut
  {NoStop}%
\bibitem [{\citenamefont {Boyarsky}\ \emph {et~al.}(2021)\citenamefont
  {Boyarsky}, \citenamefont {Ovchynnikov}, \citenamefont {Ruchayskiy},\ and\
  \citenamefont {Syvolap}}]{Boyarsky:2020dzc}%
  \BibitemOpen
  \bibfield  {author} {\bibinfo {author} {\bibfnamefont {A.}~\bibnamefont
  {Boyarsky}}, \bibinfo {author} {\bibfnamefont {M.}~\bibnamefont
  {Ovchynnikov}}, \bibinfo {author} {\bibfnamefont {O.}~\bibnamefont
  {Ruchayskiy}}, \ and\ \bibinfo {author} {\bibfnamefont {V.}~\bibnamefont
  {Syvolap}},\ }\href {\doibase 10.1103/PhysRevD.104.023517} {\bibfield
  {journal} {\bibinfo  {journal} {Phys. Rev. D}\ }\textbf {\bibinfo {volume}
  {104}},\ \bibinfo {pages} {023517} (\bibinfo {year} {2021})},\ \Eprint
  {http://arxiv.org/abs/2008.00749} {arXiv:2008.00749 [hep-ph]} \BibitemShut
  {NoStop}%
\bibitem [{\citenamefont {Chen}\ and\ \citenamefont
  {Zhang}(2025)}]{Chen:2024cla}%
  \BibitemOpen
  \bibfield  {author} {\bibinfo {author} {\bibfnamefont {Y.-M.}\ \bibnamefont
  {Chen}}\ and\ \bibinfo {author} {\bibfnamefont {Y.}~\bibnamefont {Zhang}},\
  }\href {\doibase 10.1103/741s-211w} {\bibfield  {journal} {\bibinfo
  {journal} {Phys. Rev. D}\ }\textbf {\bibinfo {volume} {111}},\ \bibinfo
  {pages} {123024} (\bibinfo {year} {2025})},\ \Eprint
  {http://arxiv.org/abs/2410.07343} {arXiv:2410.07343 [hep-ph]} \BibitemShut
  {NoStop}%
\bibitem [{\citenamefont {Ovchynnikov}(2021)}]{Ovchynnikov:2021zyo}%
  \BibitemOpen
  \bibfield  {author} {\bibinfo {author} {\bibfnamefont {M.}~\bibnamefont
  {Ovchynnikov}},\ }\emph {\bibinfo {title} {{Searches for new physics in the
  laboratory and in space}}},\ \href@noop {} {Ph.D. thesis},\ \bibinfo
  {school} {Leiden U.} (\bibinfo {year} {2021})\BibitemShut {NoStop}%
\bibitem [{\citenamefont {Aghanim}\ \emph {et~al.}(2020)\citenamefont {Aghanim}
  \emph {et~al.}}]{Planck:2018vyg}%
  \BibitemOpen
  \bibfield  {author} {\bibinfo {author} {\bibfnamefont {N.}~\bibnamefont
  {Aghanim}} \emph {et~al.} (\bibinfo {collaboration} {Planck}),\ }\href
  {\doibase 10.1051/0004-6361/201833910} {\bibfield  {journal} {\bibinfo
  {journal} {Astron. Astrophys.}\ }\textbf {\bibinfo {volume} {641}},\ \bibinfo
  {pages} {A6} (\bibinfo {year} {2020})},\ \bibinfo {note} {[Erratum:
  Astron.Astrophys. 652, C4 (2021)]},\ \Eprint
  {http://arxiv.org/abs/1807.06209} {arXiv:1807.06209 [astro-ph.CO]}
  \BibitemShut {NoStop}%
\bibitem [{\citenamefont {Sabti}\ \emph {et~al.}(2020)\citenamefont {Sabti},
  \citenamefont {Magalich},\ and\ \citenamefont {Filimonova}}]{Sabti:2020yrt}%
  \BibitemOpen
  \bibfield  {author} {\bibinfo {author} {\bibfnamefont {N.}~\bibnamefont
  {Sabti}}, \bibinfo {author} {\bibfnamefont {A.}~\bibnamefont {Magalich}}, \
  and\ \bibinfo {author} {\bibfnamefont {A.}~\bibnamefont {Filimonova}},\
  }\href {\doibase 10.1088/1475-7516/2020/11/056} {\bibfield  {journal}
  {\bibinfo  {journal} {JCAP}\ }\textbf {\bibinfo {volume} {11}},\ \bibinfo
  {pages} {056} (\bibinfo {year} {2020})},\ \Eprint
  {http://arxiv.org/abs/2006.07387} {arXiv:2006.07387 [hep-ph]} \BibitemShut
  {NoStop}%
\bibitem [{\citenamefont {Bondarenko}\ \emph {et~al.}(2018)\citenamefont
  {Bondarenko}, \citenamefont {Boyarsky}, \citenamefont {Gorbunov},\ and\
  \citenamefont {Ruchayskiy}}]{Bondarenko:2018ptm}%
  \BibitemOpen
  \bibfield  {author} {\bibinfo {author} {\bibfnamefont {K.}~\bibnamefont
  {Bondarenko}}, \bibinfo {author} {\bibfnamefont {A.}~\bibnamefont
  {Boyarsky}}, \bibinfo {author} {\bibfnamefont {D.}~\bibnamefont {Gorbunov}},
  \ and\ \bibinfo {author} {\bibfnamefont {O.}~\bibnamefont {Ruchayskiy}},\
  }\href {\doibase 10.1007/JHEP11(2018)032} {\bibfield  {journal} {\bibinfo
  {journal} {JHEP}\ }\textbf {\bibinfo {volume} {11}},\ \bibinfo {pages} {032}
  (\bibinfo {year} {2018})},\ \Eprint {http://arxiv.org/abs/1805.08567}
  {arXiv:1805.08567 [hep-ph]} \BibitemShut {NoStop}%
\bibitem [{\citenamefont {Brdar}\ \emph {et~al.}(2021)\citenamefont {Brdar},
  \citenamefont {Greljo}, \citenamefont {Kopp},\ and\ \citenamefont
  {Opferkuch}}]{Brdar:2020quo}%
  \BibitemOpen
  \bibfield  {author} {\bibinfo {author} {\bibfnamefont {V.}~\bibnamefont
  {Brdar}}, \bibinfo {author} {\bibfnamefont {A.}~\bibnamefont {Greljo}},
  \bibinfo {author} {\bibfnamefont {J.}~\bibnamefont {Kopp}}, \ and\ \bibinfo
  {author} {\bibfnamefont {T.}~\bibnamefont {Opferkuch}},\ }\href {\doibase
  10.1088/1475-7516/2021/01/039} {\bibfield  {journal} {\bibinfo  {journal}
  {JCAP}\ }\textbf {\bibinfo {volume} {01}},\ \bibinfo {pages} {039} (\bibinfo
  {year} {2021})},\ \Eprint {http://arxiv.org/abs/2007.15563} {arXiv:2007.15563
  [hep-ph]} \BibitemShut {NoStop}%
\bibitem [{\citenamefont {Pantis}\ \emph {et~al.}(1996)\citenamefont {Pantis},
  \citenamefont {Simkovic}, \citenamefont {Vergados},\ and\ \citenamefont
  {Faessler}}]{Pantis:1996py}%
  \BibitemOpen
  \bibfield  {author} {\bibinfo {author} {\bibfnamefont {G.}~\bibnamefont
  {Pantis}}, \bibinfo {author} {\bibfnamefont {F.}~\bibnamefont {Simkovic}},
  \bibinfo {author} {\bibfnamefont {J.~D.}\ \bibnamefont {Vergados}}, \ and\
  \bibinfo {author} {\bibfnamefont {A.}~\bibnamefont {Faessler}},\ }\href
  {\doibase 10.1103/PhysRevC.53.695} {\bibfield  {journal} {\bibinfo  {journal}
  {Phys. Rev. C}\ }\textbf {\bibinfo {volume} {53}},\ \bibinfo {pages} {695}
  (\bibinfo {year} {1996})},\ \Eprint {http://arxiv.org/abs/nucl-th/9612036}
  {arXiv:nucl-th/9612036} \BibitemShut {NoStop}%
\bibitem [{\citenamefont {Diamond}\ \emph {et~al.}(2024)\citenamefont
  {Diamond}, \citenamefont {Fiorillo}, \citenamefont {Marques-Tavares},
  \citenamefont {Tamborra},\ and\ \citenamefont
  {Vitagliano}}]{Diamond:2023cto}%
  \BibitemOpen
  \bibfield  {author} {\bibinfo {author} {\bibfnamefont {M.}~\bibnamefont
  {Diamond}}, \bibinfo {author} {\bibfnamefont {D.~F.~G.}\ \bibnamefont
  {Fiorillo}}, \bibinfo {author} {\bibfnamefont {G.}~\bibnamefont
  {Marques-Tavares}}, \bibinfo {author} {\bibfnamefont {I.}~\bibnamefont
  {Tamborra}}, \ and\ \bibinfo {author} {\bibfnamefont {E.}~\bibnamefont
  {Vitagliano}},\ }\href {\doibase 10.1103/PhysRevLett.132.101004} {\bibfield
  {journal} {\bibinfo  {journal} {Phys. Rev. Lett.}\ }\textbf {\bibinfo
  {volume} {132}},\ \bibinfo {pages} {101004} (\bibinfo {year} {2024})},\
  \Eprint {http://arxiv.org/abs/2305.10327} {arXiv:2305.10327 [hep-ph]}
  \BibitemShut {NoStop}%
\bibitem [{\citenamefont {Fiorillo}\ \emph
  {et~al.}(2023{\natexlab{a}})\citenamefont {Fiorillo}, \citenamefont
  {Raffelt},\ and\ \citenamefont {Vitagliano}}]{Fiorillo:2022cdq}%
  \BibitemOpen
  \bibfield  {author} {\bibinfo {author} {\bibfnamefont {D.~F.~G.}\
  \bibnamefont {Fiorillo}}, \bibinfo {author} {\bibfnamefont {G.~G.}\
  \bibnamefont {Raffelt}}, \ and\ \bibinfo {author} {\bibfnamefont
  {E.}~\bibnamefont {Vitagliano}},\ }\href {\doibase
  10.1103/PhysRevLett.131.021001} {\bibfield  {journal} {\bibinfo  {journal}
  {Phys. Rev. Lett.}\ }\textbf {\bibinfo {volume} {131}},\ \bibinfo {pages}
  {021001} (\bibinfo {year} {2023}{\natexlab{a}})},\ \Eprint
  {http://arxiv.org/abs/2209.11773} {arXiv:2209.11773 [hep-ph]} \BibitemShut
  {NoStop}%
\bibitem [{\citenamefont {Akita}\ \emph {et~al.}(2022)\citenamefont {Akita},
  \citenamefont {Im},\ and\ \citenamefont {Masud}}]{Akita:2022etk}%
  \BibitemOpen
  \bibfield  {author} {\bibinfo {author} {\bibfnamefont {K.}~\bibnamefont
  {Akita}}, \bibinfo {author} {\bibfnamefont {S.~H.}\ \bibnamefont {Im}}, \
  and\ \bibinfo {author} {\bibfnamefont {M.}~\bibnamefont {Masud}},\ }\href
  {\doibase 10.1007/JHEP12(2022)050} {\bibfield  {journal} {\bibinfo  {journal}
  {JHEP}\ }\textbf {\bibinfo {volume} {12}},\ \bibinfo {pages} {050} (\bibinfo
  {year} {2022})},\ \Eprint {http://arxiv.org/abs/2206.06852} {arXiv:2206.06852
  [hep-ph]} \BibitemShut {NoStop}%
\bibitem [{\citenamefont {Liang}\ \emph {et~al.}(2025)\citenamefont {Liang},
  \citenamefont {Su}, \citenamefont {Wu},\ and\ \citenamefont
  {Zhu}}]{Liang:2024xcx}%
  \BibitemOpen
  \bibfield  {author} {\bibinfo {author} {\bibfnamefont {Z.-L.}\ \bibnamefont
  {Liang}}, \bibinfo {author} {\bibfnamefont {L.}~\bibnamefont {Su}}, \bibinfo
  {author} {\bibfnamefont {L.}~\bibnamefont {Wu}}, \ and\ \bibinfo {author}
  {\bibfnamefont {B.}~\bibnamefont {Zhu}},\ }\href {\doibase
  10.1103/PhysRevLett.134.071001} {\bibfield  {journal} {\bibinfo  {journal}
  {Phys. Rev. Lett.}\ }\textbf {\bibinfo {volume} {134}},\ \bibinfo {pages}
  {071001} (\bibinfo {year} {2025})},\ \Eprint
  {http://arxiv.org/abs/2401.11971} {arXiv:2401.11971 [hep-ph]} \BibitemShut
  {NoStop}%
\bibitem [{\citenamefont {Li}\ \emph {et~al.}(2024)\citenamefont {Li},
  \citenamefont {Beacom}, \citenamefont {Roberts},\ and\ \citenamefont
  {Capozzi}}]{Li:2023ulf}%
  \BibitemOpen
  \bibfield  {author} {\bibinfo {author} {\bibfnamefont {S.~W.}\ \bibnamefont
  {Li}}, \bibinfo {author} {\bibfnamefont {J.~F.}\ \bibnamefont {Beacom}},
  \bibinfo {author} {\bibfnamefont {L.~F.}\ \bibnamefont {Roberts}}, \ and\
  \bibinfo {author} {\bibfnamefont {F.}~\bibnamefont {Capozzi}},\ }\href
  {\doibase 10.1103/PhysRevD.109.083025} {\bibfield  {journal} {\bibinfo
  {journal} {Phys. Rev. D}\ }\textbf {\bibinfo {volume} {109}},\ \bibinfo
  {pages} {083025} (\bibinfo {year} {2024})},\ \Eprint
  {http://arxiv.org/abs/2306.08024} {arXiv:2306.08024 [astro-ph.HE]}
  \BibitemShut {NoStop}%
\bibitem [{\citenamefont {Fiorillo}\ \emph
  {et~al.}(2023{\natexlab{b}})\citenamefont {Fiorillo}, \citenamefont
  {Heinlein}, \citenamefont {Janka}, \citenamefont {Raffelt}, \citenamefont
  {Vitagliano},\ and\ \citenamefont {Bollig}}]{Fiorillo:2023frv}%
  \BibitemOpen
  \bibfield  {author} {\bibinfo {author} {\bibfnamefont {D.~F.~G.}\
  \bibnamefont {Fiorillo}}, \bibinfo {author} {\bibfnamefont {M.}~\bibnamefont
  {Heinlein}}, \bibinfo {author} {\bibfnamefont {H.-T.}\ \bibnamefont {Janka}},
  \bibinfo {author} {\bibfnamefont {G.}~\bibnamefont {Raffelt}}, \bibinfo
  {author} {\bibfnamefont {E.}~\bibnamefont {Vitagliano}}, \ and\ \bibinfo
  {author} {\bibfnamefont {R.}~\bibnamefont {Bollig}},\ }\href {\doibase
  10.1103/PhysRevD.108.083040} {\bibfield  {journal} {\bibinfo  {journal}
  {Phys. Rev. D}\ }\textbf {\bibinfo {volume} {108}},\ \bibinfo {pages}
  {083040} (\bibinfo {year} {2023}{\natexlab{b}})},\ \Eprint
  {http://arxiv.org/abs/2308.01403} {arXiv:2308.01403 [astro-ph.HE]}
  \BibitemShut {NoStop}%
\bibitem [{\citenamefont {Bollig}\ \emph {et~al.}(2021)\citenamefont {Bollig},
  \citenamefont {Yadav}, \citenamefont {Kresse}, \citenamefont {Janka},
  \citenamefont {M{\"u}ller},\ and\ \citenamefont {Heger}}]{Bollig:2020phc}%
  \BibitemOpen
  \bibfield  {author} {\bibinfo {author} {\bibfnamefont {R.}~\bibnamefont
  {Bollig}}, \bibinfo {author} {\bibfnamefont {N.}~\bibnamefont {Yadav}},
  \bibinfo {author} {\bibfnamefont {D.}~\bibnamefont {Kresse}}, \bibinfo
  {author} {\bibfnamefont {H.~T.}\ \bibnamefont {Janka}}, \bibinfo {author}
  {\bibfnamefont {B.}~\bibnamefont {M{\"u}ller}}, \ and\ \bibinfo {author}
  {\bibfnamefont {A.}~\bibnamefont {Heger}},\ }\href {\doibase
  10.3847/1538-4357/abf82e} {\bibfield  {journal} {\bibinfo  {journal}
  {Astrophys. J.}\ }\textbf {\bibinfo {volume} {915}},\ \bibinfo {pages} {28}
  (\bibinfo {year} {2021})},\ \Eprint {http://arxiv.org/abs/2010.10506}
  {arXiv:2010.10506 [astro-ph.HE]} \BibitemShut {NoStop}%
\bibitem [{\citenamefont {Nakamura}\ \emph {et~al.}(2022)\citenamefont
  {Nakamura}, \citenamefont {Takiwaki},\ and\ \citenamefont
  {Kotake}}]{Nakamura:2022zlc}%
  \BibitemOpen
  \bibfield  {author} {\bibinfo {author} {\bibfnamefont {K.}~\bibnamefont
  {Nakamura}}, \bibinfo {author} {\bibfnamefont {T.}~\bibnamefont {Takiwaki}},
  \ and\ \bibinfo {author} {\bibfnamefont {K.}~\bibnamefont {Kotake}},\ }\href
  {\doibase 10.1093/mnras/stac1586} {\bibfield  {journal} {\bibinfo  {journal}
  {Mon. Not. Roy. Astron. Soc.}\ }\textbf {\bibinfo {volume} {514}},\ \bibinfo
  {pages} {3941} (\bibinfo {year} {2022})},\ \Eprint
  {http://arxiv.org/abs/2202.06295} {arXiv:2202.06295 [astro-ph.HE]}
  \BibitemShut {NoStop}%
\bibitem [{\citenamefont {Deshpande}\ \emph {et~al.}(1991)\citenamefont
  {Deshpande}, \citenamefont {Gunion}, \citenamefont {Kayser},\ and\
  \citenamefont {Olness}}]{Deshpande:1990ip}%
  \BibitemOpen
  \bibfield  {author} {\bibinfo {author} {\bibfnamefont {N.~G.}\ \bibnamefont
  {Deshpande}}, \bibinfo {author} {\bibfnamefont {J.~F.}\ \bibnamefont
  {Gunion}}, \bibinfo {author} {\bibfnamefont {B.}~\bibnamefont {Kayser}}, \
  and\ \bibinfo {author} {\bibfnamefont {F.~I.}\ \bibnamefont {Olness}},\
  }\href {\doibase 10.1103/PhysRevD.44.837} {\bibfield  {journal} {\bibinfo
  {journal} {Phys. Rev. D}\ }\textbf {\bibinfo {volume} {44}},\ \bibinfo
  {pages} {837} (\bibinfo {year} {1991})}\BibitemShut {NoStop}%
\bibitem [{\citenamefont {Zhang}\ \emph {et~al.}(2008)\citenamefont {Zhang},
  \citenamefont {An}, \citenamefont {Ji},\ and\ \citenamefont
  {Mohapatra}}]{Zhang:2007da}%
  \BibitemOpen
  \bibfield  {author} {\bibinfo {author} {\bibfnamefont {Y.}~\bibnamefont
  {Zhang}}, \bibinfo {author} {\bibfnamefont {H.}~\bibnamefont {An}}, \bibinfo
  {author} {\bibfnamefont {X.}~\bibnamefont {Ji}}, \ and\ \bibinfo {author}
  {\bibfnamefont {R.~N.}\ \bibnamefont {Mohapatra}},\ }\href {\doibase
  10.1016/j.nuclphysb.2008.05.019} {\bibfield  {journal} {\bibinfo  {journal}
  {Nucl. Phys. B}\ }\textbf {\bibinfo {volume} {802}},\ \bibinfo {pages} {247}
  (\bibinfo {year} {2008})},\ \Eprint {http://arxiv.org/abs/0712.4218}
  {arXiv:0712.4218 [hep-ph]} \BibitemShut {NoStop}%
\bibitem [{\citenamefont {Li}\ \emph {et~al.}(2021)\citenamefont {Li},
  \citenamefont {Ramsey-Musolf},\ and\ \citenamefont {Vasquez}}]{Li:2020flq}%
  \BibitemOpen
  \bibfield  {author} {\bibinfo {author} {\bibfnamefont {G.}~\bibnamefont
  {Li}}, \bibinfo {author} {\bibfnamefont {M.}~\bibnamefont {Ramsey-Musolf}}, \
  and\ \bibinfo {author} {\bibfnamefont {J.~C.}\ \bibnamefont {Vasquez}},\
  }\href {\doibase 10.1103/PhysRevLett.126.151801} {\bibfield  {journal}
  {\bibinfo  {journal} {Phys. Rev. Lett.}\ }\textbf {\bibinfo {volume} {126}},\
  \bibinfo {pages} {151801} (\bibinfo {year} {2021})},\ \Eprint
  {http://arxiv.org/abs/2009.01257} {arXiv:2009.01257 [hep-ph]} \BibitemShut
  {NoStop}%
\bibitem [{\citenamefont {Ramsey-Musolf}\ and\ \citenamefont
  {Vasquez}(2021)}]{Ramsey-Musolf:2020ndm}%
  \BibitemOpen
  \bibfield  {author} {\bibinfo {author} {\bibfnamefont {M.~J.}\ \bibnamefont
  {Ramsey-Musolf}}\ and\ \bibinfo {author} {\bibfnamefont {J.~C.}\ \bibnamefont
  {Vasquez}},\ }\href {\doibase 10.1016/j.physletb.2021.136136} {\bibfield
  {journal} {\bibinfo  {journal} {Phys. Lett. B}\ }\textbf {\bibinfo {volume}
  {815}},\ \bibinfo {pages} {136136} (\bibinfo {year} {2021})},\ \Eprint
  {http://arxiv.org/abs/2012.02799} {arXiv:2012.02799 [hep-ph]} \BibitemShut
  {NoStop}%
\bibitem [{\citenamefont {Dekens}\ \emph {et~al.}(2021)\citenamefont {Dekens},
  \citenamefont {Andreoli}, \citenamefont {de~Vries}, \citenamefont
  {Mereghetti},\ and\ \citenamefont {Oosterhof}}]{Dekens:2021bro}%
  \BibitemOpen
  \bibfield  {author} {\bibinfo {author} {\bibfnamefont {W.}~\bibnamefont
  {Dekens}}, \bibinfo {author} {\bibfnamefont {L.}~\bibnamefont {Andreoli}},
  \bibinfo {author} {\bibfnamefont {J.}~\bibnamefont {de~Vries}}, \bibinfo
  {author} {\bibfnamefont {E.}~\bibnamefont {Mereghetti}}, \ and\ \bibinfo
  {author} {\bibfnamefont {F.}~\bibnamefont {Oosterhof}},\ }\href {\doibase
  10.1007/JHEP11(2021)127} {\bibfield  {journal} {\bibinfo  {journal} {JHEP}\
  }\textbf {\bibinfo {volume} {11}},\ \bibinfo {pages} {127} (\bibinfo {year}
  {2021})},\ \Eprint {http://arxiv.org/abs/2107.10852} {arXiv:2107.10852
  [hep-ph]} \BibitemShut {NoStop}%
\bibitem [{\citenamefont {Czakon}\ \emph {et~al.}(1999)\citenamefont {Czakon},
  \citenamefont {Gluza},\ and\ \citenamefont {Zralek}}]{Czakon:1999ga}%
  \BibitemOpen
  \bibfield  {author} {\bibinfo {author} {\bibfnamefont {M.}~\bibnamefont
  {Czakon}}, \bibinfo {author} {\bibfnamefont {J.}~\bibnamefont {Gluza}}, \
  and\ \bibinfo {author} {\bibfnamefont {M.}~\bibnamefont {Zralek}},\ }\href
  {\doibase 10.1016/S0370-2693(99)00567-5} {\bibfield  {journal} {\bibinfo
  {journal} {Phys. Lett. B}\ }\textbf {\bibinfo {volume} {458}},\ \bibinfo
  {pages} {355} (\bibinfo {year} {1999})},\ \Eprint
  {http://arxiv.org/abs/hep-ph/9904216} {arXiv:hep-ph/9904216} \BibitemShut
  {NoStop}%
\bibitem [{\citenamefont {Li}\ \emph {et~al.}(2022)\citenamefont {Li},
  \citenamefont {Ramsey-Musolf},\ and\ \citenamefont {Vasquez}}]{Li:2022cuq}%
  \BibitemOpen
  \bibfield  {author} {\bibinfo {author} {\bibfnamefont {G.}~\bibnamefont
  {Li}}, \bibinfo {author} {\bibfnamefont {M.~J.}\ \bibnamefont
  {Ramsey-Musolf}}, \ and\ \bibinfo {author} {\bibfnamefont {J.~C.}\
  \bibnamefont {Vasquez}},\ }\href {\doibase 10.1103/PhysRevD.105.115021}
  {\bibfield  {journal} {\bibinfo  {journal} {Phys. Rev. D}\ }\textbf {\bibinfo
  {volume} {105}},\ \bibinfo {pages} {115021} (\bibinfo {year} {2022})},\
  \Eprint {http://arxiv.org/abs/2202.01789} {arXiv:2202.01789 [hep-ph]}
  \BibitemShut {NoStop}%
\bibitem [{\citenamefont {Navas}\ \emph {et~al.}(2024)\citenamefont {Navas}
  \emph {et~al.}}]{ParticleDataGroup:2024cfk}%
  \BibitemOpen
  \bibfield  {author} {\bibinfo {author} {\bibfnamefont {S.}~\bibnamefont
  {Navas}} \emph {et~al.} (\bibinfo {collaboration} {Particle Data Group}),\
  }\href {\doibase 10.1103/PhysRevD.110.030001} {\bibfield  {journal} {\bibinfo
   {journal} {Phys. Rev. D}\ }\textbf {\bibinfo {volume} {110}},\ \bibinfo
  {pages} {030001} (\bibinfo {year} {2024})}\BibitemShut {NoStop}%
\bibitem [{\citenamefont {Nemev{\v{s}}ek}\ and\ \citenamefont
  {Zhang}(2024)}]{Nemevsek:2023yjl}%
  \BibitemOpen
  \bibfield  {author} {\bibinfo {author} {\bibfnamefont {M.}~\bibnamefont
  {Nemev{\v{s}}ek}}\ and\ \bibinfo {author} {\bibfnamefont {Y.}~\bibnamefont
  {Zhang}},\ }\href {\doibase 10.1103/PhysRevD.109.056021} {\bibfield
  {journal} {\bibinfo  {journal} {Phys. Rev. D}\ }\textbf {\bibinfo {volume}
  {109}},\ \bibinfo {pages} {056021} (\bibinfo {year} {2024})},\ \Eprint
  {http://arxiv.org/abs/2312.00129} {arXiv:2312.00129 [hep-ph]} \BibitemShut
  {NoStop}%
\bibitem [{\citenamefont {Aker}\ \emph {et~al.}(2025)\citenamefont {Aker} \emph
  {et~al.}}]{KATRIN:2024cdt}%
  \BibitemOpen
  \bibfield  {author} {\bibinfo {author} {\bibfnamefont {M.}~\bibnamefont
  {Aker}} \emph {et~al.} (\bibinfo {collaboration} {KATRIN}),\ }\href {\doibase
  10.1126/science.adq9592} {\bibfield  {journal} {\bibinfo  {journal}
  {Science}\ }\textbf {\bibinfo {volume} {388}},\ \bibinfo {pages} {adq9592}
  (\bibinfo {year} {2025})},\ \Eprint {http://arxiv.org/abs/2406.13516}
  {arXiv:2406.13516 [nucl-ex]} \BibitemShut {NoStop}%
\bibitem [{\citenamefont {Abdul~Karim}\ \emph {et~al.}(2025)\citenamefont
  {Abdul~Karim} \emph {et~al.}}]{DESI:2025zgx}%
  \BibitemOpen
  \bibfield  {author} {\bibinfo {author} {\bibfnamefont {M.}~\bibnamefont
  {Abdul~Karim}} \emph {et~al.} (\bibinfo {collaboration} {DESI}),\ }\href@noop
  {} {\  (\bibinfo {year} {2025})},\ \Eprint {http://arxiv.org/abs/2503.14738}
  {arXiv:2503.14738 [astro-ph.CO]} \BibitemShut {NoStop}%
\bibitem [{\citenamefont {Tello}\ \emph {et~al.}(2011)\citenamefont {Tello},
  \citenamefont {Nemevsek}, \citenamefont {Nesti}, \citenamefont {Senjanovic},\
  and\ \citenamefont {Vissani}}]{Tello:2010am}%
  \BibitemOpen
  \bibfield  {author} {\bibinfo {author} {\bibfnamefont {V.}~\bibnamefont
  {Tello}}, \bibinfo {author} {\bibfnamefont {M.}~\bibnamefont {Nemevsek}},
  \bibinfo {author} {\bibfnamefont {F.}~\bibnamefont {Nesti}}, \bibinfo
  {author} {\bibfnamefont {G.}~\bibnamefont {Senjanovic}}, \ and\ \bibinfo
  {author} {\bibfnamefont {F.}~\bibnamefont {Vissani}},\ }\href {\doibase
  10.1103/PhysRevLett.106.151801} {\bibfield  {journal} {\bibinfo  {journal}
  {Phys. Rev. Lett.}\ }\textbf {\bibinfo {volume} {106}},\ \bibinfo {pages}
  {151801} (\bibinfo {year} {2011})},\ \Eprint {http://arxiv.org/abs/1011.3522}
  {arXiv:1011.3522 [hep-ph]} \BibitemShut {NoStop}%
\bibitem [{\citenamefont {Aad}\ \emph {et~al.}(2019{\natexlab{b}})\citenamefont
  {Aad} \emph {et~al.}}]{ATLAS:2019erb}%
  \BibitemOpen
  \bibfield  {author} {\bibinfo {author} {\bibfnamefont {G.}~\bibnamefont
  {Aad}} \emph {et~al.} (\bibinfo {collaboration} {ATLAS}),\ }\href {\doibase
  10.1016/j.physletb.2019.07.016} {\bibfield  {journal} {\bibinfo  {journal}
  {Phys. Lett. B}\ }\textbf {\bibinfo {volume} {796}},\ \bibinfo {pages} {68}
  (\bibinfo {year} {2019}{\natexlab{b}})},\ \Eprint
  {http://arxiv.org/abs/1903.06248} {arXiv:1903.06248 [hep-ex]} \BibitemShut
  {NoStop}%
\bibitem [{\citenamefont {Sirunyan}\ \emph {et~al.}(2021)\citenamefont
  {Sirunyan} \emph {et~al.}}]{CMS:2021ctt}%
  \BibitemOpen
  \bibfield  {author} {\bibinfo {author} {\bibfnamefont {A.~M.}\ \bibnamefont
  {Sirunyan}} \emph {et~al.} (\bibinfo {collaboration} {CMS}),\ }\href
  {\doibase 10.1007/JHEP07(2021)208} {\bibfield  {journal} {\bibinfo  {journal}
  {JHEP}\ }\textbf {\bibinfo {volume} {07}},\ \bibinfo {pages} {208} (\bibinfo
  {year} {2021})},\ \Eprint {http://arxiv.org/abs/2103.02708} {arXiv:2103.02708
  [hep-ex]} \BibitemShut {NoStop}%
\bibitem [{\citenamefont {Lichtenstein}\ \emph {et~al.}(2025)\citenamefont
  {Lichtenstein}, \citenamefont {Silva}, \citenamefont {Neves},\ and\
  \citenamefont {Queiroz}}]{Lichtenstein:2025pxs}%
  \BibitemOpen
  \bibfield  {author} {\bibinfo {author} {\bibfnamefont {G.}~\bibnamefont
  {Lichtenstein}}, \bibinfo {author} {\bibfnamefont {R.~C.}\ \bibnamefont
  {Silva}}, \bibinfo {author} {\bibfnamefont {M.~J.}\ \bibnamefont {Neves}}, \
  and\ \bibinfo {author} {\bibfnamefont {F.}~\bibnamefont {Queiroz}},\
  }\href@noop {} {\  (\bibinfo {year} {2025})},\ \Eprint
  {http://arxiv.org/abs/2507.10684} {arXiv:2507.10684 [hep-ph]} \BibitemShut
  {NoStop}%
\bibitem [{\citenamefont {Kriewald}\ \emph {et~al.}(2024)\citenamefont
  {Kriewald}, \citenamefont {Nemev\v{s}ek},\ and\ \citenamefont
  {Nesti}}]{Kriewald:2024cgr}%
  \BibitemOpen
  \bibfield  {author} {\bibinfo {author} {\bibfnamefont {J.}~\bibnamefont
  {Kriewald}}, \bibinfo {author} {\bibfnamefont {M.}~\bibnamefont
  {Nemev\v{s}ek}}, \ and\ \bibinfo {author} {\bibfnamefont {F.}~\bibnamefont
  {Nesti}},\ }\href {\doibase 10.1140/epjc/s10052-024-13614-8} {\bibfield
  {journal} {\bibinfo  {journal} {Eur. Phys. J. C}\ }\textbf {\bibinfo {volume}
  {84}},\ \bibinfo {pages} {1306} (\bibinfo {year} {2024})},\ \Eprint
  {http://arxiv.org/abs/2403.07756} {arXiv:2403.07756 [hep-ph]} \BibitemShut
  {NoStop}%
\bibitem [{\citenamefont {Giunti}\ and\ \citenamefont
  {Kim}(2007)}]{Giunti:2007ry}%
  \BibitemOpen
  \bibfield  {author} {\bibinfo {author} {\bibfnamefont {C.}~\bibnamefont
  {Giunti}}\ and\ \bibinfo {author} {\bibfnamefont {C.~W.}\ \bibnamefont
  {Kim}},\ }\href@noop {} {\emph {\bibinfo {title} {{Fundamentals of Neutrino
  Physics and Astrophysics}}}}\ (\bibinfo {year} {2007})\BibitemShut {NoStop}%
\bibitem [{\citenamefont {Guo}\ \emph {et~al.}(2020)\citenamefont {Guo},
  \citenamefont {Mart{\'\i}nez-Pinedo}, \citenamefont {Lohs},\ and\
  \citenamefont {Fischer}}]{Guo:2020tgx}%
  \BibitemOpen
  \bibfield  {author} {\bibinfo {author} {\bibfnamefont {G.}~\bibnamefont
  {Guo}}, \bibinfo {author} {\bibfnamefont {G.}~\bibnamefont
  {Mart{\'\i}nez-Pinedo}}, \bibinfo {author} {\bibfnamefont {A.}~\bibnamefont
  {Lohs}}, \ and\ \bibinfo {author} {\bibfnamefont {T.}~\bibnamefont
  {Fischer}},\ }\href {\doibase 10.1103/PhysRevD.102.023037} {\bibfield
  {journal} {\bibinfo  {journal} {Phys. Rev. D}\ }\textbf {\bibinfo {volume}
  {102}},\ \bibinfo {pages} {023037} (\bibinfo {year} {2020})},\ \Eprint
  {http://arxiv.org/abs/2006.12051} {arXiv:2006.12051 [hep-ph]} \BibitemShut
  {NoStop}%
\bibitem [{\citenamefont {Hannestad}\ and\ \citenamefont
  {Madsen}(1995)}]{Hannestad:1995rs}%
  \BibitemOpen
  \bibfield  {author} {\bibinfo {author} {\bibfnamefont {S.}~\bibnamefont
  {Hannestad}}\ and\ \bibinfo {author} {\bibfnamefont {J.}~\bibnamefont
  {Madsen}},\ }\href {\doibase 10.1103/PhysRevD.52.1764} {\bibfield  {journal}
  {\bibinfo  {journal} {Phys. Rev. D}\ }\textbf {\bibinfo {volume} {52}},\
  \bibinfo {pages} {1764} (\bibinfo {year} {1995})},\ \Eprint
  {http://arxiv.org/abs/astro-ph/9506015} {arXiv:astro-ph/9506015} \BibitemShut
  {NoStop}%
\bibitem [{\citenamefont {Hempel}(2015)}]{Hempel:2014ssa}%
  \BibitemOpen
  \bibfield  {author} {\bibinfo {author} {\bibfnamefont {M.}~\bibnamefont
  {Hempel}},\ }\href {\doibase 10.1103/PhysRevC.91.055807} {\bibfield
  {journal} {\bibinfo  {journal} {Phys. Rev. C}\ }\textbf {\bibinfo {volume}
  {91}},\ \bibinfo {pages} {055807} (\bibinfo {year} {2015})},\ \Eprint
  {http://arxiv.org/abs/1410.6337} {arXiv:1410.6337 [nucl-th]} \BibitemShut
  {NoStop}%
\bibitem [{\citenamefont {Hall}\ \emph {et~al.}(2010)\citenamefont {Hall},
  \citenamefont {Jedamzik}, \citenamefont {March-Russell},\ and\ \citenamefont
  {West}}]{Hall:2009bx}%
  \BibitemOpen
  \bibfield  {author} {\bibinfo {author} {\bibfnamefont {L.~J.}\ \bibnamefont
  {Hall}}, \bibinfo {author} {\bibfnamefont {K.}~\bibnamefont {Jedamzik}},
  \bibinfo {author} {\bibfnamefont {J.}~\bibnamefont {March-Russell}}, \ and\
  \bibinfo {author} {\bibfnamefont {S.~M.}\ \bibnamefont {West}},\ }\href
  {\doibase 10.1007/JHEP03(2010)080} {\bibfield  {journal} {\bibinfo  {journal}
  {JHEP}\ }\textbf {\bibinfo {volume} {03}},\ \bibinfo {pages} {080} (\bibinfo
  {year} {2010})},\ \Eprint {http://arxiv.org/abs/0911.1120} {arXiv:0911.1120
  [hep-ph]} \BibitemShut {NoStop}%
\bibitem [{\citenamefont {Shrock}(1982)}]{Shrock:1982sc}%
  \BibitemOpen
  \bibfield  {author} {\bibinfo {author} {\bibfnamefont {R.~E.}\ \bibnamefont
  {Shrock}},\ }\href {\doibase 10.1016/0550-3213(82)90273-5} {\bibfield
  {journal} {\bibinfo  {journal} {Nucl. Phys. B}\ }\textbf {\bibinfo {volume}
  {206}},\ \bibinfo {pages} {359} (\bibinfo {year} {1982})}\BibitemShut
  {NoStop}%
\bibitem [{\citenamefont {Brdar}\ \emph {et~al.}(2023)\citenamefont {Brdar},
  \citenamefont {de~Gouv\^ea}, \citenamefont {Li},\ and\ \citenamefont
  {Machado}}]{Brdar:2023tmi}%
  \BibitemOpen
  \bibfield  {author} {\bibinfo {author} {\bibfnamefont {V.}~\bibnamefont
  {Brdar}}, \bibinfo {author} {\bibfnamefont {A.}~\bibnamefont {de~Gouv\^ea}},
  \bibinfo {author} {\bibfnamefont {Y.-Y.}\ \bibnamefont {Li}}, \ and\ \bibinfo
  {author} {\bibfnamefont {P.~A.~N.}\ \bibnamefont {Machado}},\ }\href
  {\doibase 10.1103/PhysRevD.107.073005} {\bibfield  {journal} {\bibinfo
  {journal} {Phys. Rev. D}\ }\textbf {\bibinfo {volume} {107}},\ \bibinfo
  {pages} {073005} (\bibinfo {year} {2023})},\ \Eprint
  {http://arxiv.org/abs/2302.10965} {arXiv:2302.10965 [hep-ph]} \BibitemShut
  {NoStop}%
\bibitem [{\citenamefont {Atwood}\ \emph {et~al.}(2009)\citenamefont {Atwood}
  \emph {et~al.}}]{Fermilat}%
  \BibitemOpen
  \bibfield  {author} {\bibinfo {author} {\bibfnamefont {W.~B.}\ \bibnamefont
  {Atwood}} \emph {et~al.} (\bibinfo {collaboration} {Fermi-LAT}),\ }\href
  {\doibase 10.1088/0004-637X/697/2/1071} {\bibfield  {journal} {\bibinfo
  {journal} {Astrophys. J.}\ }\textbf {\bibinfo {volume} {697}},\ \bibinfo
  {pages} {1071} (\bibinfo {year} {2009})},\ \Eprint
  {http://arxiv.org/abs/0902.1089} {arXiv:0902.1089 [astro-ph.IM]} \BibitemShut
  {NoStop}%
\bibitem [{\citenamefont {Tavani}\ \emph {et~al.}(2018)\citenamefont {Tavani}
  \emph {et~al.}}]{e-ASTROGAM:2017pxr}%
  \BibitemOpen
  \bibfield  {author} {\bibinfo {author} {\bibfnamefont {M.}~\bibnamefont
  {Tavani}} \emph {et~al.} (\bibinfo {collaboration} {e-ASTROGAM}),\ }\href
  {\doibase 10.1016/j.jheap.2018.07.001} {\bibfield  {journal} {\bibinfo
  {journal} {JHEAp}\ }\textbf {\bibinfo {volume} {19}},\ \bibinfo {pages} {1}
  (\bibinfo {year} {2018})},\ \Eprint {http://arxiv.org/abs/1711.01265}
  {arXiv:1711.01265 [astro-ph.HE]} \BibitemShut {NoStop}%
\bibitem [{\citenamefont {Moiseev}\ \emph {et~al.}(2015)\citenamefont {Moiseev}
  \emph {et~al.}}]{Moiseev:2015lva}%
  \BibitemOpen
  \bibfield  {author} {\bibinfo {author} {\bibfnamefont {A.~A.}\ \bibnamefont
  {Moiseev}} \emph {et~al.},\ }\href@noop {} {\  (\bibinfo {year} {2015})},\
  \Eprint {http://arxiv.org/abs/1508.07349} {arXiv:1508.07349 [astro-ph.IM]}
  \BibitemShut {NoStop}%
\bibitem [{\citenamefont {Wu}\ \emph {et~al.}(2014)\citenamefont {Wu},
  \citenamefont {Su}, \citenamefont {Bravar}, \citenamefont {Chang},
  \citenamefont {Fan}, \citenamefont {Pohl},\ and\ \citenamefont
  {Walter}}]{Wu:2014tya}%
  \BibitemOpen
  \bibfield  {author} {\bibinfo {author} {\bibfnamefont {X.}~\bibnamefont
  {Wu}}, \bibinfo {author} {\bibfnamefont {M.}~\bibnamefont {Su}}, \bibinfo
  {author} {\bibfnamefont {A.}~\bibnamefont {Bravar}}, \bibinfo {author}
  {\bibfnamefont {J.}~\bibnamefont {Chang}}, \bibinfo {author} {\bibfnamefont
  {Y.}~\bibnamefont {Fan}}, \bibinfo {author} {\bibfnamefont {M.}~\bibnamefont
  {Pohl}}, \ and\ \bibinfo {author} {\bibfnamefont {R.}~\bibnamefont
  {Walter}},\ }\href {\doibase 10.1117/12.2057251} {\bibfield  {journal}
  {\bibinfo  {journal} {Proc. SPIE Int. Soc. Opt. Eng.}\ }\textbf {\bibinfo
  {volume} {9144}},\ \bibinfo {pages} {91440F} (\bibinfo {year} {2014})},\
  \Eprint {http://arxiv.org/abs/1407.0710} {arXiv:1407.0710 [astro-ph.IM]}
  \BibitemShut {NoStop}%
\bibitem [{\citenamefont {Jaeckel}\ \emph {et~al.}(2018)\citenamefont
  {Jaeckel}, \citenamefont {Malta},\ and\ \citenamefont
  {Redondo}}]{Jaeckel:2017tud}%
  \BibitemOpen
  \bibfield  {author} {\bibinfo {author} {\bibfnamefont {J.}~\bibnamefont
  {Jaeckel}}, \bibinfo {author} {\bibfnamefont {P.~C.}\ \bibnamefont {Malta}},
  \ and\ \bibinfo {author} {\bibfnamefont {J.}~\bibnamefont {Redondo}},\ }\href
  {\doibase 10.1103/PhysRevD.98.055032} {\bibfield  {journal} {\bibinfo
  {journal} {Phys. Rev. D}\ }\textbf {\bibinfo {volume} {98}},\ \bibinfo
  {pages} {055032} (\bibinfo {year} {2018})},\ \Eprint
  {http://arxiv.org/abs/1702.02964} {arXiv:1702.02964 [hep-ph]} \BibitemShut
  {NoStop}%
\bibitem [{\citenamefont {Prezeau}\ \emph {et~al.}(2003)\citenamefont
  {Prezeau}, \citenamefont {Ramsey-Musolf},\ and\ \citenamefont
  {Vogel}}]{Prezeau:2003xn}%
  \BibitemOpen
  \bibfield  {author} {\bibinfo {author} {\bibfnamefont {G.}~\bibnamefont
  {Prezeau}}, \bibinfo {author} {\bibfnamefont {M.}~\bibnamefont
  {Ramsey-Musolf}}, \ and\ \bibinfo {author} {\bibfnamefont {P.}~\bibnamefont
  {Vogel}},\ }\href {\doibase 10.1103/PhysRevD.68.034016} {\bibfield  {journal}
  {\bibinfo  {journal} {Phys. Rev. D}\ }\textbf {\bibinfo {volume} {68}},\
  \bibinfo {pages} {034016} (\bibinfo {year} {2003})},\ \Eprint
  {http://arxiv.org/abs/hep-ph/0303205} {arXiv:hep-ph/0303205} \BibitemShut
  {NoStop}%
\end{thebibliography}%


\end{document}